\documentclass[prd,preprint,tightenlines,floatfix,showpacs,preprintnumbers,nofootinbib]{revtex4}
\usepackage[dvips,final]{graphicx}
\usepackage{epsfig}
\usepackage{bm}
\begin{document}
\preprint{\hbox{RUB-TPII-09/03}}
\date{\today}
\title{The gluon content of the $\eta$ and
       $\eta^{\prime}$ mesons and the
       $\eta\gamma$, $\eta^{\prime}\gamma$
       electromagnetic transition form factors}
\author{\textbf{S.~S.~Agaev}}
\email{agaev_shahin@yahoo.com}
\affiliation{High Energy Physics Lab.,
             Baku State University,
             Z.\ Khalilov St.\ 23, 370148 Baku, Azerbaijan}
\author{\textbf{N.~G.~Stefanis}}
\email{stefanis@tp2.ruhr-uni-bochum.de}
\affiliation{Institut f\"ur Theoretische Physik II,
             Ruhr-Universit\"at Bochum,
             D-44780 Bochum, Germany}
\begin{abstract}
We compute power-suppressed corrections to the $\eta\gamma$ and
$\eta^{\prime}\gamma$ transition form factors
$Q^2F_{\eta(\eta^{\prime})\gamma}(Q^2)$ arising from the end-point
regions $x \to 0,\,1$ by employing the infrared-renormalon
approach. The contribution to the form factors from the quark and
gluon content of the $\eta ,\,\eta^{\prime}$ mesons is taken into
account using for the $\eta -\eta^{\prime}$ mixing the $SU_f(3)$
singlet $\eta_1$ and octet $\eta_8$ basis. The theoretical
predictions obtained this way are compared with the corresponding
CLEO data and restrictions on the input parameters (Gegenbauer
coefficients) $B_2^q(\eta_1)$, $B_2^g(\eta_1)$, and
$B_2^q(\eta_8)$ in the distribution amplitudes for the
$\eta_1,\,\,\eta_8$ states with one nonasymptotic term are
deduced. Comparison is made with the results from QCD\
perturbation theory.
\end{abstract}

\vspace{0.5 cm}
\hspace{3 cm}

\maketitle
\newpage
\section{Introduction}
\label{sec:intro} The electromagnetic transition form factors (FFs)
$F_{\rm M\gamma }(Q^2)$ of light pseudoscalar
$M\equiv\pi^0,\,\eta ,\,\,\eta^{\prime}$
mesons were the subject of much theoretical
\cite{Aga01,AM01,CCHM98,Kroll,SSK99,KP02,Kho99,SY99,BMS03,Mam03}
and experimental \cite{CLEO} research in recent years.
For instance, the CLEO collaboration reported about rather precise
measurements of the $\pi^0\gamma ,\;\eta \gamma$ and
$\eta^{\prime}\gamma$ transition FFs that stimulated interesting
theoretical investigations aiming to account for the obtained
experimental data within the framework of QCD.
One of the key objectives of such analyses is to model the
$\pi^0,\,\,\eta$ and $\eta^{\prime}$ mesons distribution amplitudes
(DAs) and, in the $\eta ,\,\,\eta^{\prime}$ case, to extract
information on their gluon components.

Indeed, it is known that the physical $\eta$ and $\eta^{\prime}$ mesons
can be represented as superpositions of a flavor $SU_f(3)$ singlet
$\eta_1$ and octet $\eta_8$ state
$$
|\eta \rangle =\cos \theta_p|\eta_8\rangle -\sin
\theta_p|\eta_1\rangle\,  ,
$$
\begin{equation}
\label{eq:1}|\eta^{\prime }\rangle =\sin \theta_p|\eta_8\rangle
+\cos \theta_p|\eta_1\rangle \, .
\end{equation}
Unlike the octet $\eta_8$ state, the $SU_f(3)$ singlet $\eta_1$
contains a two-gluon component \cite{Chase}, which even absent at
the normalization point $\mu_0^2$ , appears in the region $Q^2 >
\mu_0^2$ owing to quark-gluon mixing and renormalization-group
evolution of the $\eta_1$ state DA. The $\eta$ and $\eta^{\prime}$
mesons (cf.\ Eq.\ (\ref{eq:1})) receive a gluon contribution due
to the gluon content of the $\eta_1$ state. Because the
meson-photon transition at leading order (LO) is a pure
electromagnetic process, the gluon components of the $\eta$ and
$\eta^{\prime}$ mesons can contribute directly to the $\eta\gamma$
and $\eta^{\prime}\gamma$ transitions only at next-to-leading
order (NLO) due to quark box diagrams. They also affect the LO
result through evolution of the quark component of the $\eta$,
$\eta^{\prime}$ meson DAs. Contributions to the $\eta\gamma$ and
$\eta^{\prime}\gamma$ transition FFs, originating from the gluon
content of the $\eta$ and $\eta^{\prime}$ mesons, were recently
computed \cite{KP02} within the framework of the standard
hard-scattering approach (HSA) of the perturbative QCD (PQCD) and
estimates of the expansion parameters in the meson DAs were given.

The gluon contributions to the $\eta\gamma$ and
$\eta^{\prime}\gamma$ electromagnetic transition FFs\ are
subdominant. But in some exclusive processes, like the $B$ meson
two-body nonleptonic exclusive and semi-inclusive decays, which
involve the $\eta$ and $\eta^{\prime}$ mesons, their gluon
contribution can potentially play an essential role in explaining
the experimental data (see Ref.\ \cite{AS02} and references cited
therein). The reason is that in these processes the gluon
component of the $\eta$ and $\;\eta^{\prime}$ mesons contributes
to corresponding hard-scattering amplitudes already at LO of
perturbative QCD. Hence, the gluonic parts of the
$\eta,\;\;\eta^{\prime}$ meson DAs, deduced from the
$\eta\gamma,\;\eta^{\prime}\gamma$ data, are important input
ingredients in studying a wide range of exclusive processes, given
that they are universal, i.e., process- and frame-independent
quantities.

The HSA\ and the perturbative QCD factorization theorems
\cite{ERBL}, at asymptotically large values of the momentum
transfer $Q^2$, lead to reliable predictions for exclusive
processes. But in the momentum-transfer regime of a few GeV${}^2$,
experimentally accessible at present for most exclusive processes,
power-suppressed corrections $(1/Q^2)^n,\;n=1,2,3\ldots $ may play
an important role in explaining the experimental data. In order to
evaluate such corrections, the QCD\ running-coupling (RC) method,
combined with the infrared (IR) renormalon approach, was proposed
\cite{Aga01,AM01,AS02,Aga95a,Aga95b}. This method allows one to
evaluate power-behaved contributions in exclusive processes
arising from the end-point regions $x\to 0,\;1$. In this manner,
power corrections to the electromagnetic FFs\ $F_{\rm M}(Q^2)$
(${\rm M}\equiv \pi$, K) \cite{Aga95a,Aga95b}, to the transition
FFs\ $F_{\rm M\gamma}(Q^2)$ (${\rm M}\equiv
\pi^0,\;\eta,\;\eta^{\prime}$) \cite{Aga01,AM01}, as well as to
the gluon-gluon-$\eta^{\prime}$ vertex function \cite{AS02} were
computed. Power corrections can also be obtained by means of the
Landau-pole free expression for the QCD coupling constant
\cite{SS97}. This analytic approach was used to calculate in a
unifying way power corrections to the electromagnetic pion FF and
such to the inclusive cross section of the Drell-Yan process
\cite{KS01,Ste02}.

Power corrections to the $\eta\gamma$ and $\eta^{\prime}\gamma$
electromagnetic transition FFs\ within the RC\ method were computed
in Ref.\ \cite{Aga01} and predictions for the structure of the DAs
of the $\eta$ and $\eta^{\prime}$ mesons were made.
In the present work we extend this sort of investigation by also
including into the calculation of the
$\eta \gamma,\;\eta^{\prime}\gamma$ transition FFs\ the power
corrections originating from the gluonic content of the
$\eta,\,\,\eta^{\prime}$ mesons that were not taken into account in
Ref.\ \cite{Aga01}.
This will enable us to extract their DAs from comparing our
theoretical predictions with the CLEO data \cite{CLEO}.

The paper is structured as follows: Sec.\ II contains the required
information on the hard-scattering amplitudes for the
$\eta_1\gamma$ and $\eta_8\gamma$ transitions, accounting also for
the gluon content of the $\eta_1$ state. The DAs of the $SU_f(3)$
singlet $\eta_1$ and octet $\eta_8$ states are considered and
their evolution is taken into account. In Sec.\ III we compute the
$\eta\gamma$ and $\eta^{\prime}\gamma$ transition FFs\ within the
RC\ method and obtain the Borel resummed expressions for them. The
asymptotic limit $Q^2\rightarrow\infty$ of these FFs\ is explored
and the standard HSA\ leading-twist predictions for the FFs\ are
recovered. In Sec.\ IV we perform a numerical analysis and compare
our results with the CLEO \cite{CLEO} data with the aim to extract
constraints on the $\eta$ and $\eta^{\prime}$ meson DAs. Finally,
Sec.\ V contains our concluding remarks.

\section{$\mathbf{SU_f(3)}$ singlet and octet components of the
         $\mathbf{\eta\gamma}$, $\mathbf{\eta^{\prime}\gamma}$
         transition Form factors}
\label{sec:SU3f}
The meson-photon electromagnetic transition FF $F_{\rm M\gamma}(Q^2)$
can be defined in terms of the invariant amplitude $\Gamma^\mu$ of
the process\footnote{Hereafter M denotes the $\eta$ or $\eta^{\prime}$
meson.}
\begin{equation}
\label{eq:2}\gamma^{*}(q_1)+\gamma (q_2)\rightarrow M(p)
\end{equation}
in the following way
\begin{equation}
\label{eq:3}\Gamma^\mu = ie^2F_{\rm M\gamma}(Q^2)\epsilon^{\mu \nu
\alpha \beta}\epsilon_\nu (q_2)q_{1\alpha}q_{2\beta}\, ,
\end{equation}
where $\epsilon_\nu (q_2)$ is the polarization vector of the real
photon and $Q^2=-q_1^2$. The
FFs\ of the $\eta\gamma$ and $\eta^{\prime}\gamma$ transitions are
sums of the corresponding singlet $F_{\rm M\gamma}^1(Q^2)$ and
octet $F_{\rm M\gamma}^8(Q^2)$ contributions
\begin{equation}
\label{eq:4}F_{\rm M\gamma}(Q^2)=F_{\rm M\gamma}^1(Q^2)+F_{\rm
M\gamma}^8(Q^2)\,.
\end{equation}
The FF\ of the octet state $F_{\rm M\gamma}^8(Q^2)$ and the
quark-related component of the FF of the singlet state, $F_{\rm
M\gamma }^1(Q^2)$, can be computed by employing the results
obtained for the pion-photon transition FF \cite{AC81,KP02}. The
latter is known at $O(\alpha_{{\rm s}})$ of PQCD\ \cite{AC81}.
More recently, also a part of $O(\alpha_{{\rm s}}^2)$ corrections
were computed \cite{MNP01}. The gluonic component of the singlet
contribution $F_{\rm M\gamma}^1(Q^2)$ was just recently calculated
within the framework of the HSA\ of the perturbative QCD in Ref.\
\cite{KP02}.

In accordance with the PQCD\ factorization theorems, at large
momentum-transfer, the FFs\ $F_{\rm M\gamma }^1(Q^2)$ and $F_{\rm
M\gamma }^8(Q^2)$ can be represented in the form of a convolution
of the corresponding hard-scattering amplitudes with the quark and
gluon components of the DAs of the $\eta_1$ and $\eta_8$ states,
$$
Q^2F_{\rm M\gamma}^1(Q^2)=f_{\rm M}^1N_1\left\{ T_{{\rm H},\,
0}^q(x) \otimes \phi_{\eta_1}^q(x,\mu_{\rm F}^2)+\right.
$$
\begin{equation}
\label{eq:5}
\frac{\alpha_{{\rm s}}(\mu_{\rm R}^2)}{4\pi} \, C_{\rm
F}\left. \left[ T_{{\rm H},1}^q(x,Q^2,\mu_{\rm F}^2)\otimes
\phi_{\eta_1}^q(x,\mu_{\rm F}^2) +T_{{\rm H},1}^g(x,Q^2,\mu_{\rm
F}^2)\otimes \phi_{\eta_1}^g(x,\mu_{\rm F}^2)\right] \right\}
\end{equation}
and
\begin{equation}
\label{eq:6}
Q^2F_{\rm M\gamma }^8(Q^2)=f_{\rm M}^8N_8\left[
T_{{\rm H},\,0}^q(x) \otimes \phi_{\eta_8}(x,\mu_{\rm
F}^2)+\frac{\alpha_{{\rm s}}(\mu_{\rm R}^2)}{4\pi} \, C_{\rm F}
T_{{\rm H},1}^q(x,Q^2,\mu_{\rm F}^2)\otimes
\phi_{\eta_8}(x,\mu_{\rm F}^2)\right] \, ,
\end{equation}
where all quantities above are renormalized, i.e., are
singularity-free, and the symbol $\otimes$ denotes the convolution
$$
T_{\rm H}(x)\otimes \phi (x)=\int_0^1dxT_{\rm H}(x)\phi (x)\, .
$$
Here the functions $T_{{\rm H},\,0}^q(x)$ and
$T_{{\rm H},1}^q(x,Q^2,\mu_{\rm F}^2)$ are the hard-scattering
amplitudes for the partonic subprocess
$\gamma + \gamma^{*} \to q + \overline{q}$
at LO\ and NLO, respectively, and
$T_{{\rm H},1}^g(x,Q^2,\mu _{\rm F}^2)$
is the NLO hard-scattering amplitude for
$\gamma + \gamma^{*} \to g+g$, with
$\mu_{\rm F}^2$, $\mu_{\rm R}^2$
being the factorization and renormalization scales.
In Eqs.\ (\ref{eq:5}) and (\ref{eq:6}), $f_{\rm M}^i$ are the
M-meson decay constants, $C_{\rm F}=4/3$ is the color factor, and
$N_1$ and $N_8$ are numerical constants, each depending on the
quark structure of the associated $\eta_1$, $\eta_8$ states
\begin{equation}
\label{eq:8}
N_1=\frac 1{\sqrt{3}}\left( e_u^2+e_d^2+e_s^2\right)
,\;\;\;\;\; N_8=\frac 1{\sqrt{6}}\left(
e_u^2+e_d^2-2e_s^2\right)\, .
\end{equation}

The hard-scattering amplitudes $T_{{\rm H},\,0}^q(x)$, $T_{{\rm
H},1}^q(x,Q^2,\mu _{\rm F}^2)$, and $T_{{\rm H},1}^g(x,Q^2,\mu
_{\rm F}^2)$ are well-known \cite{KP02,AC81,MNP01} and are given
by the following expressions

$$
T_{{\rm H},\,0}^q(x)=\frac 1x+\frac 1{\overline{x}}\; ,
$$
$$
T_{{\rm H},1}^q(x,Q^2,\mu_{\rm F}^2) =\frac 1x\left[ \ln
{}^2x-\frac{x\ln x}{\overline{x}} -9\right] +\frac 1x(3+2\ln x)\ln
\frac{Q^2}{\mu_{\rm F}^2}+(x\leftrightarrow \overline{x})\, ,
$$
\begin{equation}
\label{eq:9}
T_{{\rm H},1}^g(x,Q^2,\mu_{\rm F}^2) =\frac{x\ln
{}^2x}{\overline{x}}+\left( 6-\frac 4{\overline{x}}\right) \ln
{}x+2\frac{x\ln x}{\overline{x}}\ln \frac{Q^2}{\mu_{\rm F}^2}
-(x\leftrightarrow \overline{x})\, ,
\end{equation}
where $\overline{x}\equiv 1-x$.

The next ingredients needed for computing the FFs\ $F_{\rm
M\gamma}^1(Q^2)$ and $F_{\rm M\gamma}^8(Q^2)$ are the meson-decay
constants $f_{\rm M}^i$ and the distribution amplitudes
$\phi_{\eta_1}^q(x,\mu_{\rm F}^2),\;\phi_{\eta_1}^g(x,\mu_{\rm
F}^2)$, and $\phi_{\eta_8}(x,\mu_{\rm F}^2)$ for the
$\eta_1,\;\eta_8$ states.
The decay constants $f_{\rm M}^i$ are defined as the matrix
elements of the axial-vector currents
$J_{\mu 5}^i$ with $i=1,8$
\begin{equation}
\label{eq:10}<0|J_{\mu 5}^i|M(p)>=if_{\rm M}^ip_\mu \;.
\end{equation}
In the octet-singlet basis the constants $f_{\rm M}^i$ can be
parameterized by two methods.
One is to follow the pattern of state mixing (cf.\ Eq.\ \ref{eq:1}))
$$
f_\eta ^8=f_8\cos \theta_p\, ,\;\;f_\eta^1=-f_1\sin \theta_p \, ,
$$
\begin{equation}
\label{eq:11}f_{\eta^{\prime}}^8 =f_8\sin \theta_p\,
,\;\;f_{\eta^{\prime}}^1 =f_1\cos \theta_p \, ,
\end{equation}
where the decay constants $f_1, \,\,f_8$ and $\theta_p$ are given
by \cite{FKS98}
\begin{equation}
\label{eq:12}f_1=1.17f_\pi \, ,\,\,\,f_8=1.26f_\pi\,
,\,\,\,\theta_p =-15.4^{\circ}
\end{equation}
with $f_\pi =0.131\,\,{\rm GeV}$ being the pion decay constant.

The second method employs a two-mixing-angles parametrization:
$$
f_\eta^8=f_8\cos \theta_8\, ,\;f_\eta^1=-f_1\sin \theta_1\, ,
$$
\begin{equation}
\label{eq:13}
f_{\eta^{\prime}}^8=f_8\sin \theta_8\,
,\,\,\,f_{\eta^{\prime}}^1 =f_1\cos \theta_1
\end{equation}
with the mixing angles $\theta_1$ and $\theta_8$ provided by
\cite{FKS98}
\begin{equation}
\label{eq:14}
\theta_1=-9.2^{\circ}\, ,\;\theta_8=-21.2^{\circ}\, .
\end{equation}

Parametrization (\ref{eq:11}) leads to simple expressions for the
physical FFs in terms of $F_{\eta_1 \gamma}(Q^2)$ and
$F_{\eta_8\gamma}(Q^2)$; viz.,

$$
F_{\eta\gamma}(Q^2)=F_{\eta_8\gamma}(Q^2)\cos\theta_p-
F_{\eta_1\gamma}(Q^2)\sin\theta_p,
$$

\begin{equation}
\label{eq:15}
F_{\eta^{\prime}\gamma}(Q^2)=F_{\eta_8\gamma}(Q^2)\sin\theta_p+
F_{\eta_1\gamma}(Q^2)\cos\theta_p \, ,
\end{equation}
where the form factors $F_{\eta_1\gamma}(Q^2)$ and
$F_{\eta_8\gamma}(Q^2)$ are determined by expressions (\ref{eq:5})
and (\ref{eq:6}), but with the decay constants $f_M^i$ replaced by
$f_i$. In our numerical computations we shall use both schemes the
conventional one-angle mixing scheme and also the
two-mixing-angles parametrization.

The main question still to be answered concerns the shape of the
DAs of the $\eta_1$ and $\eta_8$ states. In general, a meson DA\
is a function containing all nonperturbative, long-distance
effects, which cannot be calculated by employing perturbative QCD\
methods. Nonetheless, as a direct consequence of factorization,
the evolution of the DA\ with the factorization scale $\mu_{\rm
F}^2$ is governed by PQCD. Input information at the starting point
of evolution, i.e., the dependence of the DA\ on the variable $x$
at the normalization point $\mu_0^2$, has to be extracted from
experimental data or derived via nonperturbative methods, for
example, QCD\ sum rules with nonlocal condensates \cite{BMS01}
(see also \cite{BM02}) or instanton-based models \cite{inst} at
some (low) momentum scale, characteristic of the particular
nonperturbative model.

Because of mixing of the quark-antiquark component with the
two-gluon part of the DA, the evolution equation for the DA\ of
the flavor singlet pseudoscalar $\eta_1$ state has a $2\times 2$
matrix form \cite{Chase}. The solution of this equation is given
by the expressions

\begin{equation}
\label{eq:16}
\phi^q(x,\mu_{\rm F}^2) = 6x\overline{x}\left(
1+\sum_{n=2,4..}^\infty
\left\{ B_n^q\left[ \frac{\alpha_{{\rm s}}(\mu_0^2)}{%
\alpha_{{\rm s}}(\mu_{\rm F}^2)}%
            \right]^{\frac{\gamma_{+}^n}{\beta_0}}+
\rho_n^gB_n^g\left[ \frac{\alpha_{{\rm s}}(\mu_0^2)}{\alpha_{{\rm
s}}(\mu_{\rm F}^2)}
             \right]^{\frac{\gamma_{-}^n}{\beta_0}}
\right\} C_n^{3/2}(x-\overline{x})\right)
\end{equation}
and
\begin{equation}
\label{eq:17}
\phi ^g(x,\mu_{\rm F}^2) =
x\overline{x}\sum_{n=2,4..}^\infty \left\{ \rho_n^qB_n^q\left[
\frac{\alpha_{{\rm s}}(\mu_0^2)}{\alpha_{{\rm s}}(\mu_{\rm F}^2)}
\right]^{\frac{\gamma_{+}^n}{\beta_0}}+ B_n^g\left[
\frac{\alpha_{{\rm s}}(\mu_0^2)}{\alpha_{{\rm s}}(\mu_{\rm F}^2)}
\right]^{\frac{\gamma_{-}^n}{\beta_0}}\right\}
C_{n-1}^{5/2}(x-\overline{x})\, .
\end{equation}
Here $C_n^{3/2}(z)$ and $C_n^{5/2}(z)$ are Gegenbauer polynomials.
Detailed information concerning the parameters
$\rho_n^q,\,\,\rho_n^g$ and the anomalous dimensions
$\gamma_{+}^n,\,\,\gamma_{-}^n$ can be found in Ref.\ \cite{AS02}.
In Eqs.\ (\ref{eq:16}) and (\ref{eq:17}) the coefficients $B_n^q$
and $B_n^g$ will be considered as free input parameters, the
values of which at the normalization point $\mu_0^2$ determine the
shapes of the DAs $\phi^q(x,\mu_{\rm F}^2)$ and
$\phi^g(x,\mu_{\rm F}^2)$.

In our calculations we shall use a phenomenological DA for the
$\eta_1$ state containing only the first Gegenbauer polynomials
$C_2^{3/2}(x-\overline{x})$ and $C_1^{5/2}(x-\overline{x})$ (i.e.,
$B_2^q\neq 0,\,\,B_2^g\neq 0$ and $B_n^q=B_n^g=0$ for all $n>2$)
\begin{equation}
\label{eq:18}
C_2^{3/2}(x-\overline{x})=6(1-5x\overline{x}),\,\,\,\,%
\,C_1^{5/2}(x-\overline{x})=5(x-\overline{x})\, .
\end{equation}
Under this assumption, the DAs\ assume the following forms
\cite{AS02}
$$
\phi _{\eta _1}^q(x,\mu_{\rm F}^2) =6x\overline{x}\left[
1+A(\mu_{\rm F}^2)- 5A(\mu_{\rm F}^2)x\overline{x} \right] \, ,
$$
\begin{equation}
\label{eq:19}
\phi _{\eta _1}^g(x,\mu_{\rm
F}^2)=x\overline{x}(x-\overline{x})B(\mu_{\rm F}^2)\, .
\end{equation}
For $n_f=3$, in other words, for momentum transfers $ Q^2$ below
the charm-quark production threshold, the functions $A(\mu_{\rm
F}^2)$ and $B(\mu_{\rm F}^2)$ are defined by
$$
A(\mu_{\rm F}^2) =6B_2^q\left[ \frac{\alpha_{{\rm s}}(\mu_{\rm
F}^2)}{\alpha_{{\rm s}}(\mu_0^2)}
\right]^{\frac{48}{81}}-\frac{B_2^g}{15} \left[\frac{\alpha _{{\rm
s}}(\mu_{\rm F}^2) }{\alpha_{{\rm s}}(\mu_0^2)}
\right]^{\frac{101}{81}},
$$
\begin{equation}
\label{eq:20}
B(\mu_{\rm F}^2)
=16B_2^q\left[ \frac{\alpha_{{\rm s}}(\mu_{\rm F}^2)}{\alpha_{%
{\rm s}}(\mu_0^2)}\right]^{\frac{48}{81}}
+5B_2^g\left[\frac{\alpha_{{\rm s}}(\mu_{\rm F}^2)}{\alpha_{{\rm
s}}(\mu_0^2)} \right]^{\frac{101}{81}}.
\end{equation}

The DA of the octet $\eta_8$ state contains only the quark
component $\phi_{\eta_8}(x,\mu_{\rm F}^2)$. This DA is identical
to $\phi_{\eta_1}^q(x,\mu_{\rm F}^2)$, but with $A(\mu_{\rm F}^2)$
replaced by $C(\mu_{\rm F}^2)$, i.e.,
\begin{equation}
\label{eq:21}C(\mu_{\rm F}^2)
=6B_2^q\left[ \frac{\alpha _{{\rm s}}(\mu_{\rm F}^2)}{\alpha _{{\rm %
s}}(\mu _0^2)}\right] ^{\frac{50}{81}}.
\end{equation}

The explicit expressions for the functions $A(\mu_{\rm F}^2)$ and
$B(\mu_{\rm F}^2)$ at momentum transfers above the charm quark
threshold (or, for $n_f=4$) can be found in the Appendix of Ref.\
\cite{AS02}.
For $n_f=4$, the function $C(\mu_{\rm F}^2)$ should be modified to read
$50/81 \to 2/3$.
If necessary, we shall distinguish between input parameters in
Eqs.\ (\ref{eq:20}) and (\ref{eq:21}) by using the notations
$B_2^q(\eta_1)$ and $B_2^q(\eta_8)$.

\section{Borel resummed $\mathbf{\eta\gamma}$ and
         $\,\,$ $\mathbf{\eta^{\prime}\gamma}$ transition form
         factors}
\label{sec:borel} In Sec.\ \ref{sec:SU3f} we have outlined the key
ingredients pertaining to both the standard HSA\ as well as the
RC\ treatment of the transition FFs\ $F_{\rm M\gamma}^1(Q^2)$ and
$F_{\rm M\gamma}^8(Q^2)$. Let us now turn to a discussion of the
main differences between these two approaches, starting with the
choice of the scales $\mu_{\rm R}^2$ and $\mu_{\rm F}^2$. It is
evident that if a physical quantity can be factorized, like Eqs.\
(\ref{eq:5}) and (\ref{eq:6}), then the left-hand side (LHS) \
cannot depend on artificial intrinsic scales or on the particular
renormalization and factorization schemes adopted. But at any
finite order of QCD perturbation theory, truncation of the
corresponding perturbative series will give rise to a dependence
on the scales $\mu_{\rm F}^2$ and $\mu_{\rm R}^2$, as well as on
the factorization and renormalization scheme (for an in-depth
discussion of these issues, we refer the reader to the second
paper of Ref.\ \cite{SSK99}). Because higher-order corrections in
perturbative QCD\ computations are, as a rule, large for both
inclusive and exclusive processes, reliable theoretical
predictions require an optimal scale-setting that minimizes
higher-order corrections. Typically, the factorization scale
enters the NLO\ contribution to the hard-scattering amplitude of
meson transition or electromagnetic form factors in the form $\sim
\ln (Q^2/\mu_{\rm F}^2)$, so that taking $\mu_{\rm F}^2$ equal to
$Q^2$ eliminates this term. But in order to analyze the
sensitivity of our results to a chosen value of $\mu_{\rm F}^2$,
we shall perform all analytical computations for $\mu_{\rm F}^2
\neq Q^2$.

The choice of the renormalization scale is somewhat subtler
because this scale enters not only the NLO\ contribution, but also
as the argument of the running strong coupling
$\alpha_{{\rm s}}(\mu_{\rm R}^2)$.
To discuss this question, consider first the scale of the strong
coupling.
One effective method to solve this problem is the
Brodsky-Lepage-Mackenzie (BLM) scale-setting procedure \cite{BLM83}.
In this framework, a large part of the higher-order
corrections---namely, those originating from the diagrams with quark
``bubbles'' insertions---can be absorbed into the scale of the QCD\
coupling constant.
When utilizing this new scale one finds the NLO\ correction to be
significantly reduced relative to its initial value.
The generalization of the BLM\ procedure to all orders of
perturbative QCD\ led to the invention of the RC\ method and the
IR\ renormalon approach (for a review, see Ref.\ \cite{Ben99}).
In the case of inclusive processes, it was proven by explicit
calculation that all-order resummation of diagrams with a chain of
(quark) bubble insertions into the gluon line gives the same results
as the calculation of one-loop Feynman diagrams for the quantity under
consideration using the QCD running coupling at the vertices.
Moreover, the IR\ renormalon approach in conjunction with the
``ultraviolet dominance hypothesis'' enables one to estimate
higher-twist corrections to a wide range of inclusive
processes.

This approach was used for studying IR\ renormalon effects in
exclusive processes as well.
For instance, $(-\beta_0\alpha_{{\rm s}}/4\pi )^n$ corrections to
the Brodsky-Lepage evolution kernel $V[x,y;\alpha_{{\rm s}}(Q^2)]$
were computed in Ref.\ \cite{GK98,Mik98} and renormalon-chain
contributions to the pseudoscalar meson DA and the $\pi^0\gamma$
transition FF were taken into account in \cite{GK98}.
Similar investigations along this line of thought were performed
in Refs.\ \cite{BC98,ASS98}.

In addition to loop-integration ambiguities, exclusive processes
may receive power-behaved contributions from the end point regions
due to the integration in a process amplitude over the
longitudinal momentum fractions of the involved partons. In fact,
in order to reduce the NLO correction, for example, to the pion
electromagnetic FF $F_\pi(Q^2)$, the renormalization scale
$\mu_{\rm R}^2$ should be set equal to the typical four-momentum,
flowing through hard gluon lines in the partonic subprocess
$q\overline{q}^{\prime}+\gamma^{*}\rightarrow
q\overline{q}^{\prime}$ \cite{BLM83}. Choosing the scale $\mu_{\rm
R}^2$ this way, inevitably leads to a dependence on the
longitudinal momentum fractions carried by the hadron's
constituents. In the case of $F_\pi(Q^2)$, the NLO contribution to
the hard-scattering amplitude $T_{\rm H}^1(x,y,Q^2)$ contains a
logarithm of the form
$\ln\left(\overline{x}\overline{y}Q^2/\mu_{\rm R}^2\right)$, with
$x$ and $y$ being, respectively, the longitudinal momentum
fractions of the quarks in the initial and the final pion. Hence,
the natural choice to eliminate this term would be to set
$\mu_{\rm R}^2=\overline{x}\overline{y}Q^2$. But due to the
convolution of the hard-scattering with the soft components (cf.\
e.g., Eq.\ (\ref{eq:5})), integrations over $x$ and $y$ appear
that give rise to power corrections when approaching the end point
$x\rightarrow 0,1;\,\,y\rightarrow 0,1$ regions. Renormalizing the
process amplitude at a scale close to the large external momentum
$Q^2$ makes such contributions less pronounced but at the expense
of large NLO logarithms. Therefore, if we are to optimize our
theoretical calculation, we have to minimize NLO contributions
while keeping under control power corrections in the end point
regions.

Specifically, for the meson-photon transition we have
\begin{equation}
\label{eq:21b}
\mu_{\rm R}^2=Q^2x\, ,\;\; \;\;\overline{\mu}_{\rm
R}^2=Q^2\overline{x}\, ,
\end{equation}
because in the corresponding two, leading-order, Feynman diagrams
the absolute values of the square of the momenta flowing through
virtual quark lines are determined exactly by these expressions.
In the standard HSA\ one ``freezes'' the scale of the QCD coupling
constant $\mu_{\rm R}^2$ ($\overline{\mu}_{\rm R}^2$), by
replacing $x$ by its mean value $<x>=1/2$ and performs then the
integrations in Eqs.\ (\ref{eq:5}) and (\ref{eq:6}) with
$\alpha_{{\rm s}}(Q^2/2)$.
Let us stay within the HSA and concentrate on the NLO corrections
to the quark component of Eq.\ (\ref{eq:5}).
Omitting an unimportant, in the present context, constant factor,
we get
$$
Q^2F_{\rm M\gamma}^1(Q^2)_1^{\rm quark}\sim \alpha_{{\rm
s}}\left(\frac{Q^2}2 \right)t(x,\mu_{\rm F}^2) \otimes
\phi_{\eta_1}^q(x,\mu_{\rm F}^2)+\alpha_{{\rm
s}}\left(\frac{Q^2}2\right)t(\overline{x},\mu_{\rm F}^2) \otimes
\phi_{\eta_1}^q(x,\mu_{\rm F}^2)=
$$
\begin{equation}
\label{eq:22}
2\alpha_{{\rm s}}\left(\frac{Q^2}2
\right)t(x,\mu_{\rm F}^2)\otimes \phi_{\eta_1}^q(x,\mu_{\rm
F}^2)\, ,
\end{equation}
where the function $t(x,\mu_{\rm F}^2)$ is
\begin{equation}
\label{eq:23}
t(x,\mu_{\rm F}^2)=\frac 1x\left[ \ln
{}^2x-\frac{x\ln x}{\overline{x}} +\left(3+2 \ln x \right)a
-9\right]
\end{equation}
and $a=\ln (Q^2 / \mu_{\rm F}^2)$.
In deriving Eq.\ (\ref{eq:22}) we used the symmetry property of
the quark component of the $\eta_1$ state DA, valid also for the
function $\phi_{\eta_8}(x,\mu_{\rm F}^2)$,
\begin{equation}
\label{eq:24}
\phi _{\eta _1}^q(x,\mu_{\rm F}^2)=\phi _{\eta
_1}^q(\overline{x} , \mu_{\rm F}^2)\, ,\;\;\phi _{\eta
_8}(x,\mu_{\rm F}^2)=\phi _{\eta _8}(\overline{x},\mu_{\rm F}^2)\,.
\end{equation}
The generalization of our analysis to encompass the gluon component
of the FF is straightforward.

Applying the RC\ method, the same quark component of the
$\eta_1\gamma$ transition FF takes the form

$$
Q^2F_{\rm M\gamma}^1(Q^2)_1^{\rm quark}\sim \alpha_{{\rm
s}}(Q^2x)t(x,\mu_{\rm F}^2)\otimes \phi_{\eta_1}^q(x,\mu_{\rm
F}^2)+\alpha_{{\rm s}}(Q^2\overline{x})t(\overline{x},\mu_{\rm
F}^2) \otimes \phi_{\eta_1}^q(x,\mu_{\rm F}^2)=
$$
\begin{equation}
\label{eq:25}2\alpha_{{\rm s}}(Q^2x)t(x,\mu_{\rm F}^2)\otimes
\phi_{\eta_1}^q(x,\mu_{\rm F}^2)\, .
\end{equation}

After a similar analysis for the gluon component of the form
factor $Q^2F_{\rm M\gamma}^1(Q^2)$, using the RC\ method, we find
\begin{equation}
\label{eq:27} Q^2F_{\rm M\gamma}^1(Q^2)_1^{\rm gluon}\sim
2\alpha_{{\rm s}}(Q^2x)g(x,\mu_{\rm F}^2) \otimes
\phi_{\eta_1}^g(x,\mu_{\rm F}^2)\, ,
\end{equation}
with the function
$g(x,\mu_{\rm F}^2)$ being given by the expression
\begin{equation}
\label{eq:28}g(x,\mu_{\rm F}^2)=\frac{x\ln {}^2x}{\overline{x}}
+\left( 6-\frac 4{\overline{x}}\right) \ln x +2a\frac{x \ln
x}{\overline{x}}
\end{equation}
by making use of the antisymmetry of the gluon DA\
$\phi_{\eta_1}^g(x,\mu_{\rm F}^2)$ under the exchange
$x\leftrightarrow \overline{x}$
\begin{equation}
\label{eq:29}\phi_{\eta_1}^g(x,\mu_{\rm
F}^2)=-\phi_{\eta_1}^g(\overline{x},\mu_{\rm F}^2)\, .
\end{equation}
The gluon component in the standard HSA has the same form
(\ref{eq:27}) with the argument of $\alpha_{\rm s}$ being replaced
by $Q^2/2$.

Summing up, we can write the transition FFs\
$Q^2F_{\rm M\gamma}^1(Q^2)$ and $Q^2F_{\rm M\gamma}^8(Q^2)$
in the context of the RC\ method as follows
$$
Q^2F_{\rm M\gamma}^1(Q^2)=f_{\rm M}^1N_1\left\{ T_{{\rm H},0}^q(x)
\otimes \phi_{\eta_1}^q(x,\mu_{\rm F}^2)\right.
$$
\begin{equation}
\label{eq:30}
+ \frac{C_{\rm F}}{2\pi }\left. \left[ \alpha _{{\rm s}%
}(Q^2x)t(x,\mu_{\rm F}^2)\otimes \phi _{\eta _1}^q(x,\mu_{\rm F}^2)+
\alpha _{{\rm s}%
}(Q^2x)g(x,\mu_{\rm F}^2)\otimes \phi _{\eta _1}^g(x,\mu
_F^2)\right] \right\}
\end{equation}
and
\begin{equation}
\label{eq:31}
Q^2F_{\rm M\gamma }^8(Q^2) =f_{\rm M}^8N_8\left[
T_{{\rm H},0}^q(x)\otimes \phi_{\eta_8}(x,\mu_{\rm F}^2)
+\frac{C_{\rm F}}{2\pi}\alpha_{{\rm s}}(Q^2x)t(x,\mu_{\rm F}^2)
\otimes \phi_{\eta_8}(x,\mu_{\rm F}^2)\right] .
\end{equation}

But the integrations over $x$ in Eqs.\ (\ref{eq:30}) and
(\ref{eq:31}), when retaining the $x$ dependence of the QCD
coupling $\alpha_{\rm s}(Q^2x)$ [$\alpha_{\rm
s}(Q^2\overline{x})$], lead to divergent integrals because the
running coupling $\alpha_{\rm s}(Q^2x)$ [$\alpha_{\rm
s}(Q^2\overline{x})$] suffers from an infrared singularity in the
limit $x\to 0$ [$x\to 1$]. This means that in order to perform
calculations with the running coupling, some procedure for its
regularization in the end point $x \to 0,\;1$ regions has to be
adopted.

As a first step in this direction, we express the running coupling
$\alpha_{\rm s}(Q^2x)$ in terms of $\alpha_{\rm s}(Q^2)$, employing
the renormalization-group equation, to find \cite{CS},
\begin{equation}
\label{eq:31a}
\alpha_{\rm s}(Q^2x) \simeq \frac{\alpha_{\rm s}(Q^2)}{1+\ln x/t}
\left[1 - \frac{\alpha_{\rm s}(Q^2)\beta_1}{2\pi\beta_0}
\frac{\ln [1+ \ln x/t]}{1+ \ln x /t}\right]\, ,
\end{equation}
where $\alpha_{\rm s}(Q^2)$ is the one-loop QCD coupling,
$t=4\pi/\beta_0\alpha_{\rm s}(Q^2)=\ln (Q^2/\Lambda^2)$, with
$\beta_0$ and $\beta_1$ being the one- and two-loop coefficients of
the QCD beta function
$$
\beta_0=11-\frac2{3}\,n_f\, ,\;\;\;\beta_1=51-\frac{19}{3}n_f\, ,
$$
respectively.
Equation (\ref{eq:31a}) expresses $\alpha_{\rm s}(Q^2x)$ in terms of
$\alpha_{\rm s}(Q^2)$ to the $\sim\alpha_{\rm s}^2(Q^2)$ order
accuracy.

Inserting (\ref{eq:31a}) into the formulas for the transition FFs\
(\ref{eq:30}) and (\ref{eq:31}), we obtain integrals which are
still divergent, but can be computed using existing methods.
One of them, applied in \cite{Aga95a} for the calculation of the
electromagnetic pion form factor, allows one to obtain the
quantity under consideration as a perturbative series in
$\alpha_{\rm s}(Q^2)$ with factorially growing coefficients
$C_n \sim (n-1)!$.
Similar series may be found also for the transition FFs\
$Q^2F_{\rm M\gamma}^i(Q^2)$

\begin{equation}
\label{eq:31b} Q^2F_{\rm M\gamma}^i(Q^2) \sim \sum_{n=1}^{\infty}
\left[\frac{\alpha_{\rm s}(Q^2)} {4\pi} \right]^n
\beta_0^{n-1}C_n\, .
\end{equation}

But a perturbative QCD series with factorially growing
coefficients is a signal for the IR renormalon nature of the
divergences in (\ref{eq:31b}).
The convergence radius of the series (\ref{eq:31b}) is zero and its
resummation should be performed by employing the Borel integral
technique.
First, one has to find the Borel transform
$B[Q^2F_{\rm M\gamma}^i](u)$ of the corresponding series \cite{HZ}

\begin{equation}
\label{eq:31c}
B\left[ Q^2F_{\rm M\gamma}^i \right
](u)=\sum_{n=1}^{\infty}\frac{u^{n-1}}{(n-1)!}\, C_n\, .
\end{equation}
Because the coefficients of the series (\ref{eq:31b}) behave like
$C_n \sim (n-1)!$, the Borel transform (\ref{eq:31c}) contains
poles located on the positive $u$ axis of the Borel plane.
In other words, the divergence of the series (\ref{eq:31b}) has been
transformed into pole singularities of the function
$B[Q^2F_{\rm M\gamma}^i](u)$.
These poles are exactly the IR renormalon poles.

Now in order to define the sum (\ref{eq:31b}), or to find the
resummed expression for the form factors, one has to invert
$B[Q^2F_{\rm M\gamma}^i](u)$ to get

\begin{equation}
\label{eq:31d} \left[Q^2F_{\rm M\gamma}^i(Q^2) \right ]^{\rm res}
\sim P.V. \int_0^{\infty} du\exp\left[-\frac{4\pi u}
{\beta_0\alpha_{\rm s}(Q^2)} \right] B\left[Q^2F_{\rm M\gamma}^i
\right](u)\, ,
\end{equation}
and remove the IR renormalon divergences by the principal value
prescription.
These intermediate steps can be bypassed by introducing the inverse
Laplace transforms of the functions in (\ref{eq:31a}), i.e.,
\begin{equation}
\label{eq:31l}
\frac1{(t+z)^{\nu}}=
\frac1{\Gamma(\nu)}\int_{0}^{\infty}du
\exp[-u(t+z)]u^{\nu-1},\;\;\;Re\nu
> 0
\end{equation}
and
\begin{equation}
\label{eq:31m} \frac{\ln [t+z]}{(t+z)^2}=\int_{0}^{\infty}du \exp
[-u(t+z)](1-\gamma_E-\ln u)u\, ,
\end{equation}
where $\Gamma(z)$ is the Gamma function,
$\gamma_E \simeq 0.577216$ is the Euler constant, and
$z=\ln x$ [or z=$\ln \overline{x}$].

Then using (\ref{eq:31a}), (\ref{eq:31l}), and (\ref{eq:31m}) we
find \cite{Aga01,AS02}

\begin{equation}
\label{eq:33}\alpha_{{\rm
s}}(Q^2x)=\frac{4\pi}{\beta_0}\int_0^\infty
due^{-ut}R(u,t)x^{-u}\, .
\end{equation}
Here, the function $R(u,t)$ is defined as
\begin{equation}
\label{eq:34} R(u,t)=1-\frac{2\beta_1}{\beta_0^2}u(1-\gamma_E-\ln
t-\ln u)\, .
\end{equation}

Having used Eqs.\ (\ref{eq:19}), (\ref{eq:23}), (\ref{eq:28}), and
(\ref{eq:33}) in Eqs.\ (\ref{eq:30}) and (\ref{eq:31}), and
performing the integrations over $x$, we obtain the FFs\
$Q^2F_{\rm M\gamma}^1(Q^2)$ and $Q^2F_{\rm M\gamma}^8(Q^2)$ within
the RC\ method; viz.,
$$
Q^2F_{\rm M\gamma}^1(Q^2)=f_{\rm M}^1N_1\left\{ 6+A(\mu_{\rm
F}^2)+\frac{12C_{\rm F}}{\beta_0} \left[\left( 1+A(\mu_{\rm
F}^2)\right)\int_0^\infty due^{-ut}R(u,t)Q_1(u)-\right. \right.
$$
\begin{equation}
\label{eq:35}
\left. \left. 5A(\mu_{\rm F}^2)\int_0^\infty
due^{-ut}R(u,t)Q_2(u)\right] +\frac{2C_{\rm F}}{\beta
_0}B(\mu_{\rm
F}^2)\int_0^\infty due^{-ut}R(u,t)G(u)\right\}
\end{equation}
and
$$
Q^2F_{\rm M\gamma }^8(Q^2) =f_{\rm M}^8N_8\left\{ 6+C(\mu_{\rm
F}^2)+\frac{12C_{\rm F}}{\beta _0}\left[ \left( 1+C(\mu_{\rm
F}^2)\right) \int_0^\infty due^{-ut}R(u,t)Q_1(u)-\right. \right.
$$
\begin{equation}
\label{eq:36}
\left. \left. 5C(\mu_{\rm F}^2)\int_0^\infty
due^{-ut}R(u,t)Q_2(u)\right] \right\} .
\end{equation}
The functions $Q_1(u),\,\,Q_2(u)$, and $G(u)$ have the expressions
\begin{equation}
\label{eq:37}
Q_1(u)=\frac {d^2}{d\beta^2}B(2,\beta)_{1-u}-
\frac{d}{d\beta}B(1,\beta)_{2-u} +2a
\frac{d}{d\beta}B(2,\beta)_{1-u}+3(a-3)B(1-u,2)\, ,
\end{equation}

\begin{equation}
\label{eq:38}
Q_2(u)=\frac{d^2}{d\beta^2}B(3,\beta)_{2-u}
-\frac{d}{d\beta}B(2,\beta)_{3-u}+
2a\frac{d}{d\beta}B(3,\beta)_{2-u}+3(a-3)B(2-u,3)\, ,
\end{equation}
and
$$
G(u)=\frac{d^2}{d\beta^2}B(1,\beta)_{4-u}+
6\frac{d}{d\beta}B(2,\beta)_{3-u}-
4\frac{d}{d\beta}B(1,\beta)_{3-u}+2a\frac{d}{d\beta}B(1,\beta)_{4-u}-
$$
\begin{equation}
\label{eq:39}
\frac{d^2}{d\beta^2}B(2,\beta)_{3-u}-6\frac{d}{d\beta}B(3,\beta)_{2-u}+
4\frac{d}{d\beta}B(2,\beta)_{2-u}-2a\frac{d}{d\beta}B(2,\beta)_{3-u}\,
,
\end{equation}
where the standard notation for the Beta function
$B(x,y)$
$$
B(x,y)=\frac{\Gamma (x)\Gamma (y)}{\Gamma (x+y)}
$$
has been employed.

After some manipulations, the functions $Q_1(u),\,\,Q_2(u)$ and
$G(u)$ can be recast into the more convenient forms
$$
Q_1(u)=\frac 2{(1-u)^3}-\frac 2{(2-u)^3}-\frac{2a}{(1-u)^2}+\frac
{1+2a}{(2-u)^2}+3\frac{a-3}{(1-u)(2-u)}\, ,
$$
$$
Q_2(u)=\frac 2{(2-u)^3}-\frac 4{(3-u)^3}+
\frac 2{(4-u)^3}-\frac{2a}{(2-u)^2}+
$$
$$
\frac {1+4a}{(3-u)^2}-\frac
{1+2a}{(4-u)^2}+6\frac{a-3}{(2-u)(3-u)(4-u)}\, ,
$$
\begin{equation}
\label{eq:42} G(u)=\frac 4{(4-u)^3}-\frac 2{(3-u)^3}+\frac
2{(2-u)^2} -2\frac{5-a}{(3-u)^2}+4\frac{3-a}{(4-u)^2}\, .
\end{equation}

One observes that the FFs\ given by (\ref{eq:35}) and
(\ref{eq:36}) contain a finite number of single, double, and
triple poles located at the points $u_0=1,\,2,\,3\,,\,4$. In other
words, employing expression (\ref{eq:33}), we have transformed the
end point $x\to 0$ singularities in Eqs.\ (\ref{eq:30}) and
(\ref{eq:31}) into (multiple) poles in the Borel plane $u$. These
poles are the IR renormalon poles and consequently the integrals
in Eqs.\ (\ref{eq:35}) and (\ref{eq:36}) are just the inverse
Borel transformations (\ref{eq:31d}), in which the Borel
transforms $B[Q^2F_{\rm M\gamma}^{1(8)}](u)$ of the NLO parts of
the quark components and the gluon component of the scaled FFs
are, up to constant factors, proportional to the functions
$$
R(u,t)Q_1(u),\; R(u,t)Q_2(u),\; R(u,t)G(u)\, .
$$

As we have emphasized above, the IR renormalon divergences can be
cured by employing the principal value prescription, which we
adopt in this work to regularize divergent integrals.
Therefore, the integrals over $u$ in Eqs.\ (\ref{eq:35}) and
(\ref{eq:36}) are to be understood in the sense of the Cauchy
principal value.
Removing these divergences, Eqs.\ (\ref{eq:35}) and (\ref{eq:36})
become just the Borel resummed expressions $[Q^2F_{\rm
M\gamma}^1(Q^2)]^{\rm res}$ and
$[Q^2F_{\rm M\gamma}^8(Q^2)]^{\rm res}$ for these scaled FFs.

It is known \cite{Aga01,AS02} that the IR renormalon pole located
at the point $u_0=n$ of the Borel plane corresponds to the
power-suppressed correction $\sim (1/Q^2)^n$, contained in the
scaled FFs. To make the discussion of this question as transparent
as possible, let us for the time being neglect the nonleading term
$\sim \alpha_{\rm s}^2$ in (\ref{eq:31a}) and make the replacement
$R(u,t) \to 1$ in (\ref{eq:33}). Then, the integrals in the
resummed FFs\ with multiple IR renormalon poles at $u_0=n$ can be
easily expressed in terms of the integrals with a single IR
renormalon pole at the same point $u_0=n$ (see, Eq.\ (\ref{eq:47})
below), so that our formulas (\ref{eq:35}) and (\ref{eq:36}) will
contain the integrals
\begin{equation}
\label{eq:42a}
\frac{4\pi}{\beta_0}\int_0^{\infty}\frac{e^{-ut}du}{n-u}=
\int_0^1\alpha_{\rm s}(Q^2x)x^{n-1}dx=\frac1{n}f_{2n}(Q)\, ,
\end{equation}
where $f_{2n}(Q)$ are the moment integrals
\begin{equation}
\label{eq:42b} f_p(Q)=\frac{p}{Q^p}\int_0^{Q}dkk^{p-1}\alpha_{\rm
s}(k^2)\, .
\end{equation}
The integrals $f_p(Q)$ were calculated before \cite{WEB} using the
IR matching scheme:
\begin{equation}
\label{eq:42c} f_p(Q)=\left(\frac{\mu_{I}}{Q} \right)^pf_p(\mu_I)+
\alpha_{\rm s}(Q^2)\sum_{n=0}^{N}\left [ \frac{\beta_0}{2\pi
p}\,\alpha_{\rm s}(Q^2)\right ]^n[n!-\Gamma(n+1,\;p\ln
(Q/\mu_I))]\, ,
\end{equation}
where $\mu_I$ is the infrared matching scale and $\Gamma(n+1,\;z)$
is the incomplete Gamma function. In Eq.\ (\ref{eq:42c})
$\{f_p(\mu_I)\}$ are phenomenological parameters, representing the
weighted average of $\alpha_{\rm s}(k^2)$ over the IR region $0 <k
<\mu_I$, and act at the same time as infrared regulators of the
right-hand side (RHS)\ of Eq.\ (\ref{eq:42a}). The first term on
the RHS\ of Eq.\ (\ref{eq:42c}) is a power-suppressed contribution
to $f_p(Q)$ and models the ``soft'' part of the moment integral.
It cannot be calculated within PQCD, whereas the second term on
the RHS\ of Eq.\ (\ref{eq:42c}) is the perturbatively calculable
part of the function $f_p(Q)$, representing its ``hard''
perturbative tail. In other words, the infrared matching scheme
allows one to estimate power corrections to the moment integrals
by explicitly pulling them out from the full expression,
 and introducing new
nonperturbative parameters $f_p(\mu_I)$. The same moment integrals
$f_{2n}(Q)$, computed in the framework of the RC\ method (LHS\ of
Eq.\ (\ref{eq:42a})), contain information on both their soft and
the perturbative components. Indeed, numerical calculations
demonstrate that the LHS\ of (\ref{eq:42a}), computed by employing
the principal value prescription, and its RHS---found by means of
(\ref{eq:42c}) for $p \geq 2$---practically coincide with each
other. Therefore, we can state that the scaled and resummed FFs\
(\ref{eq:35}) and (\ref{eq:36}) contain power corrections $\sim
(1/Q^2)^n,\; n=1,2,3,4$. Hence the usage in phenomenological
applications of both the IR matching scheme and the RC\ method
seems legitimate. In fact, both methods have been used to
calculate the pion's electromagnetic FF \cite{Aga99} and the
vertex function $Q^2F_{\eta^{\prime}g^{*}g^{*}}(Q^2,\omega)$
\cite{AS02}. But the RC\ method has an advantage over the IR
matching scheme because it allows one to compute the functions
$f_p(Q)$ without introducing the new nonperturbative parameters
$\mu_I$ and $f_p(\mu_I)$. Moreover, using this method, the
parameters $f_p(\mu_I)$ themselves can be computed in good
agreement with model calculations and available experimental data
\cite{Aga99,AS02}.

The power corrections $\sim (1/Q^2)^n$ are important in the region
of moderate $Q^2$ and change the behavior of the scaled and
resummed FFs\ (\ref{eq:35}) and (\ref{eq:36}) as functions of
$Q^2$ significantly, both qualitatively and quantitatively. In the
present work we have to deal only with a finite number of IR
renormalon poles. Their number and location, in the case under
consideration, depend on the DAs\ (\ref{eq:19}) used in the
calculations. The asymptotic DAs\ of the $\eta_1$ and $\eta_8$
states lead to only two IR renormalon poles at $u_0=1,\;2$.
Distribution amplitudes, which include higher-order Gegenbauer
polynomials $n>2$, may lead to a series of IR renormalon poles at
$u_0=5,\;6 \ldots$. Note that at small momentum transfers, in each
integral $f_{2n}(Q)$ associated with the pole $u_0=n$, the soft
part dominates. In the context of the IR matching scheme the
integral $f_{2n}(Q)$ at $Q^2=\mu_I^2$ even consists of just the
soft contribution. Restricting our considerations to contributions
arising only from the nearest to the origin $u=0$ IR renormalon
poles (which are, of course, the dominant ones), entails two
problems: first, it reduces the accuracy of the numerical results
and second, one loses information on the DAs\ of the $\eta_1$ and
$\eta_8$ states. Therefore, for the self-consistent treatment of
the FFs\ (\ref{eq:35}) and (\ref{eq:36}) we should take into
account contributions coming from all IR renormalon poles.

The principal value prescription, adopted here to regularize
divergent integrals over $u$, generates itself power-suppressed
ambiguities (uncertainties)
$$
\sim N_q\frac{\Phi_q(Q^2)}{Q^{2q}}\, ,
$$
where $\{\Phi_q(Q^2)\}$ are calculable functions entirely
determined by the residues of the Borel transforms
$B[Q^2F_{\rm M\gamma}^{1(8)}](u)$ at the pole $q=u_0$ and
$\{N_q\}$ are arbitrary constants.
Taking into account these ambiguities in Eqs.\ (\ref{eq:35})
and (\ref{eq:36}) leads to a modification of the Borel resummed
FFs, amounting to
\begin{equation}
\label{eq:42d}[Q^2F_{\rm M\gamma}^{1(8)}(Q^2)]^{\rm res}
\rightarrow [Q^2F_{\rm M\gamma}^{1(8)}(Q^2)]^{\rm res}
+[Q^2F_{\rm M\gamma}^{1(8)}(Q^2)]^{\rm amb}\, .
\end{equation}
In accordance with the ``ultraviolet dominance hypothesis'', the
uncertainty in Eq.\ (\ref{eq:42d}) will allow us to estimate power
corrections to the scaled FFs\ stemming from sources other than
the end point integrations. Indeed, by fitting the constants
$\{N_q\}$ to the experimental data, one can deduce some
information concerning the magnitude of such corrections.

It should be clear that regardless of the methods employed for the
computation of the form factors, in the limit
$Q^2\rightarrow\infty$ these must reach their asymptotic values.
The important problem to be clarified is then whether our resummed
expressions
$[Q^2F_{\eta \gamma}(Q^2)]^{\rm res}$,
$[Q^2F_{\eta^{\prime}\gamma}(Q^2)]^{\rm res}$ lead in the limit
$Q^2\rightarrow \infty$ to their corresponding well-known asymptotic
forms.
For the sake of simplicity, we restrict ourselves to the
$\mu_{\rm F}^2=Q^2$ case.
To answer the question posed above, we first explore the
$Q^2\rightarrow \infty$ limits of the DAs\
$\phi_{\eta_1}^q(x,Q^2)$, $\phi_{\eta_1}^g(x,Q^2)$, and
$\phi_{\eta_8}(x,Q^2)$.
Because in Eqs.\ (\ref{eq:16}) and (\ref{eq:17}) the eigenvalues
$\gamma_{\pm}^n<0$ and their absolute values increase with $n$ for
all $n\geq 2$, going to the asymptotic limit only the quark component
of the $\eta_1$ state DA survives, evolving to its asymptotic form,
whereas the DA of the gluon component $\phi_{\eta_1}^g(x,Q^2)$ in
this limit vanishes, i.e.,
$$
\phi _{\eta _1}^q(x,Q^2)
\stackrel{Q^2\rightarrow \infty }{\longrightarrow }
6x \overline{x},\; \phi _{\eta _1}^g(x,Q^2)
\stackrel{Q^2\rightarrow \infty }{%
\longrightarrow }0\, .
$$
The same arguments apply also to the DA of the $\eta_8$ state,
consisting only of the quark component
$$
\phi _{\eta _8}(x,Q^2)\stackrel{Q^2\rightarrow \infty
}{\longrightarrow }6x \overline{x}\, .
$$
In our case this means that the following limits are fulfilled
\begin{equation}
\label{eq:43}A(Q^2),\; B(Q^2),\; C(Q^2)\stackrel{Q^2\rightarrow
\infty}{\longrightarrow}0\, .
\end{equation}
Moreover, we have to take into account that in this limit the term
$\sim \alpha_{{\rm s}}^2(Q^2)$ in the expansion of
$\alpha_{{\rm s}}(Q^2x)$ in terms of $\alpha_{{\rm s}}(Q^2)$ has
to be neglected \cite{Aga01,AS02}.
The latter requirement is equivalent to the replacement
\begin{equation}
\label{eq:44}
\int_0^\infty e^{-ut}R(u,t)du\rightarrow
\int_0^\infty e^{-ut}du\, .
\end{equation}
Then we obtain
\begin{equation}
\label{eq:45}
[Q^2F_{\rm M\gamma}(Q^2)]^{\rm
res}\stackrel{Q^2\rightarrow \infty}{\longrightarrow} 6\left(
f_{\rm M}^1N_1+f_{\rm M}^8N_8\right) \left[ 1+\frac{2C_{\rm
F}}{\beta_0}\int_0^\infty du e^{-ut}Q_1(u)\right] .
\end{equation}
But this is not the final result because in the integral above
$t=\ln (Q^2/\Lambda^2)$ and its $Q^2\rightarrow \infty $ limit has
still to be computed.
The integral
\begin{equation}
\label{eq:46}\int_0^\infty due^{-ut}\left[ \frac 2{(1-u)^3}
-\frac2{(2-u)^3}+\frac 1{(2-u)^2}-\frac 9{1-u}+\frac 9{2-u}\right]
\end{equation}
can be expressed in terms of the logarithmic integral
\begin{equation}
\label{eq:48}
li(x)=P.V.\int_0^x\frac{dt}{\ln t}
\end{equation}
after performing the integration by parts of the first three terms
to obtain
\begin{equation}
\label{eq:47}
\int_0^\infty \frac{e^{-ut}du}{(n-u)^3}=-\frac
1{2n^2} -\frac{\ln \lambda}{2n} +\frac{\ln
{}^2\lambda}2\frac{li(\lambda^n)}{\lambda^n}\, , \,\,\int_0^\infty
\frac{e^{-ut}du}{(n-u)^2} =-\frac 1n+\ln {}\lambda
\frac{li(\lambda^n)}{\lambda^n}\, ,
\end{equation}
where $\lambda =Q^2/\Lambda ^2$.
Employing the formula \cite{AS02}
$$
\frac{li(x^n)}{x^n} \simeq \frac 1{n\ln
x}\sum_{m=0}^M\frac{m!}{(n\ln x)^m} \, ,\: \; \; M\gg 1
$$
and retaining in the expressions
$$
\ln{}^2\lambda\frac{li(\lambda^n)}{\lambda^n}\, ,
\,\,\ln{}\lambda\frac{li(\lambda^n)}{\lambda^n}
$$
terms up to $O(1/\ln \lambda )$ order, we finally find
$$
[Q^2F_{\rm M\gamma}(Q^2)]^{\rm res} \stackrel{Q^2\rightarrow
\infty}{\longrightarrow} 6\left( f_{\rm M}^1N_1+f_{\rm
M}^8N_8\right)\left[ 1-\frac 5{3\pi}\, \alpha_{{\rm s}
}(Q^2)\right] =
$$

\begin{equation}
\label{eq:49}
\frac 1{\sqrt{3}}\left( 4f_{\rm M}^1+\sqrt{2}f_{\rm
M}^8\right) \left[ 1-\frac 5{3\pi}\, \alpha_{{\rm s}}(Q^2) \right]
.
\end{equation}
It is worth noting that numerical constants and terms
$\sim \ln\lambda$
in Eq.\ (\ref{eq:46}), appearing due to Eq.\ (\ref{eq:47}), cancel
out in the final result.

The Eq.\ (\ref{eq:49}) for M$=\eta$ and $\eta^{\prime}$ supplies
the asymptotic limit of the corresponding transition FFs. These
limits can be readily obtained within the standard HSA\ by means
of the asymptotic DAs\ of the $\eta_1$ and $\,\,\eta_8$ states.
Stated differently, by explicit computation we have proved that in
the $Q^2\rightarrow \infty$ limit the Borel resummed expressions
(\ref{eq:35}) and (\ref{eq:36}) lead to the well-known asymptotic
forms of the $F_{\eta \gamma}(Q^2)$ and
$F_{\eta^{\prime}\gamma}(Q^2)$ form factors.

\begin{figure}[t]
\centering\epsfig{file=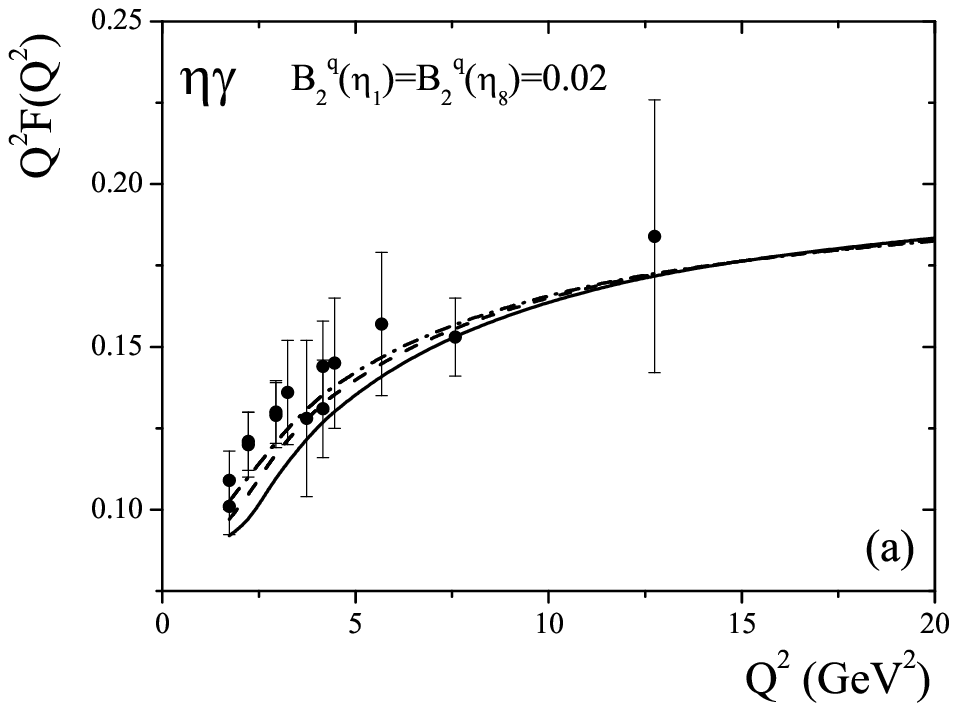,height=8cm,width=7.0cm,clip=}
\hspace{1.5cm}
\centering\epsfig{file=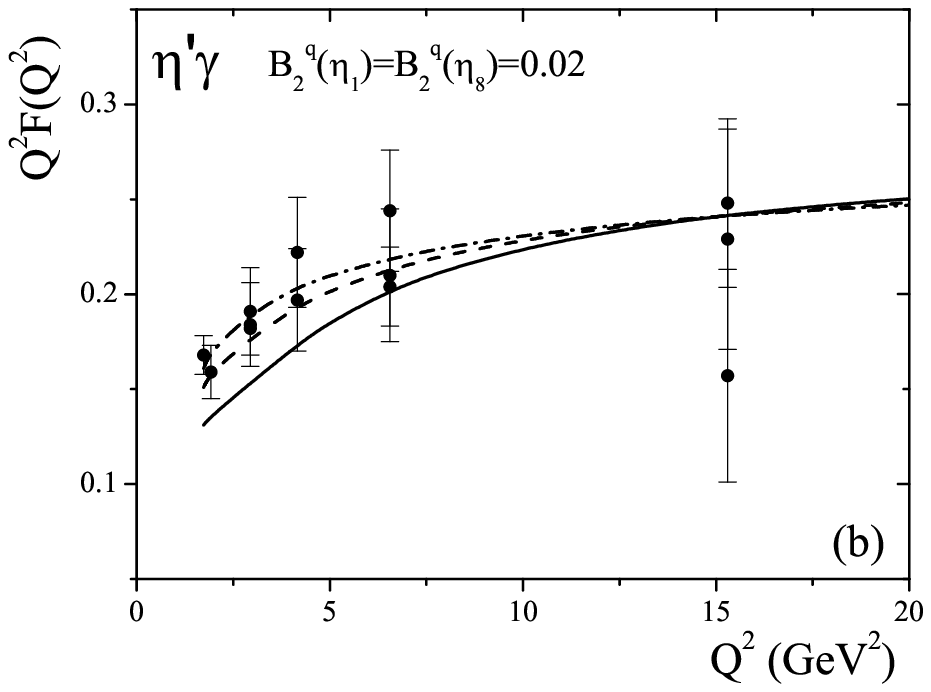,height=7.8cm,width=6.3cm,clip=}
\vspace{0.0cm} \caption{Predictions for the scaled form factors
as functions of $Q^2$ of the $\eta\gamma$ (left panel) and
$\eta^{\prime}\gamma$ (right panel) electromagnetic transition.
For the solid curves the designation is $B_2^g(\eta _1)=0$.
The dashed lines correspond to $B_2^g(\eta _1)=10$; for the
dash-dotted curves we use $B_2^g(\eta _1)=15$.
The data are taken from Ref.\ \cite{CLEO}.} \label{fig:Gfig1}
\end{figure}

\section{Extracting the $\mathbf{\eta}$ and $\mathbf{\eta^{\prime}}$
         meson Distribution Amplitudes}
\label{sec:etaeta^primeDAs} In this section we perform numerical
computations of the Borel resummed $\eta \gamma$ and
$\eta^{\prime}\gamma$ transition FFs\footnote{Notice that in this
Section ``FF'' means the scaled form factors.} in order to extract
the $\eta$ and $\eta^{\prime}$ meson DAs\ from the CLEO\ data. We
shall also compare our theoretical predictions with those obtained
with the standard HSA\ \cite{KP02,AP03}, the aim being to reveal
the role of power corrections at low-momentum-transfer in the
exclusive process under consideration. In our calculations below
we shall use the following values of $\Lambda, \;\; \mu_0^2$ and
$\mu_{\rm F}^2$
\begin{equation}
\label{eq:50}
\Lambda_4 =0.25\,{\rm GeV},\;\mu_0^2=1\,{\rm GeV}^2,
\mu_{\rm F}^2=Q^2
\end{equation}
and we shall employ both the one-angle scheme (\ref{eq:11}) and
also the two-mixing-angles scheme (\ref{eq:13}). Eqs.\
(\ref{eq:20}) and (\ref{eq:21}) will be evaluated using the
two-loop approximation for the QCD coupling $\alpha_{{\rm
s}}(Q^2)$:
\begin{equation}
\label{eq:alf}
\alpha_{{\rm s}}(Q^2)=\frac{4\pi }{\beta_0\ln (Q^2/\Lambda^2)}
\left[ 1-\frac{2\beta_1}{\beta_0^2}
\frac{\ln \ln (Q^2/\Lambda^2)}{\ln (Q^2/\Lambda ^2)}\right] .
\end{equation}

\begin{figure}[t]
\centering\epsfig{file=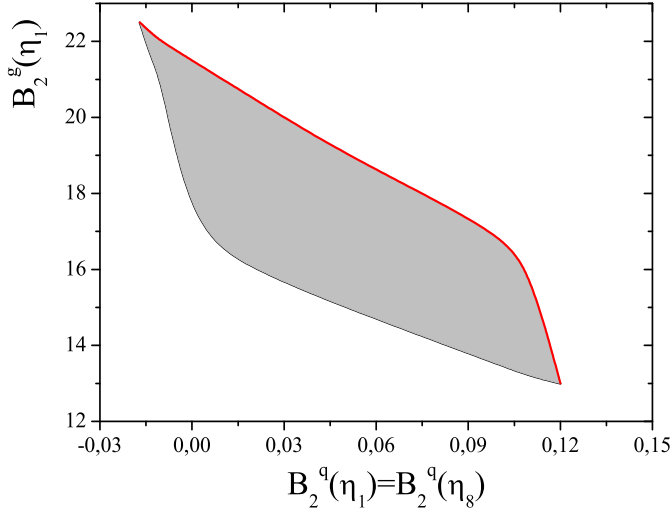,height=8.0cm,width=10.1cm,clip=}
\hspace{1.5cm}
\vspace{0.0cm} \caption{The $1\sigma$ area in the $B_2^{g}-B_2^q$
plane estimated within the RC\ method by comparing the CLEO data and
the theoretical predictions for the resummed and scaled transition
FFs\ $\eta\gamma$, $\eta^{\prime}\gamma$.}
\label{fig:Gfig2}
\end{figure}

\begin{figure}[b]
\centering\epsfig{file=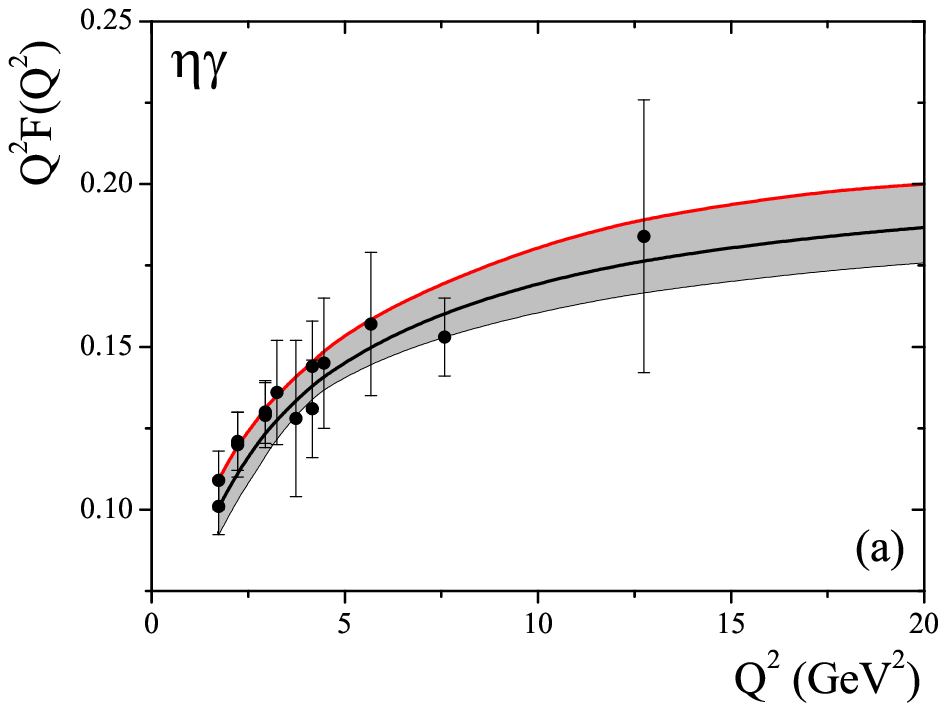,height=8cm,width=7.0cm,clip=}
\hspace{1.5cm}
\centering\epsfig{file=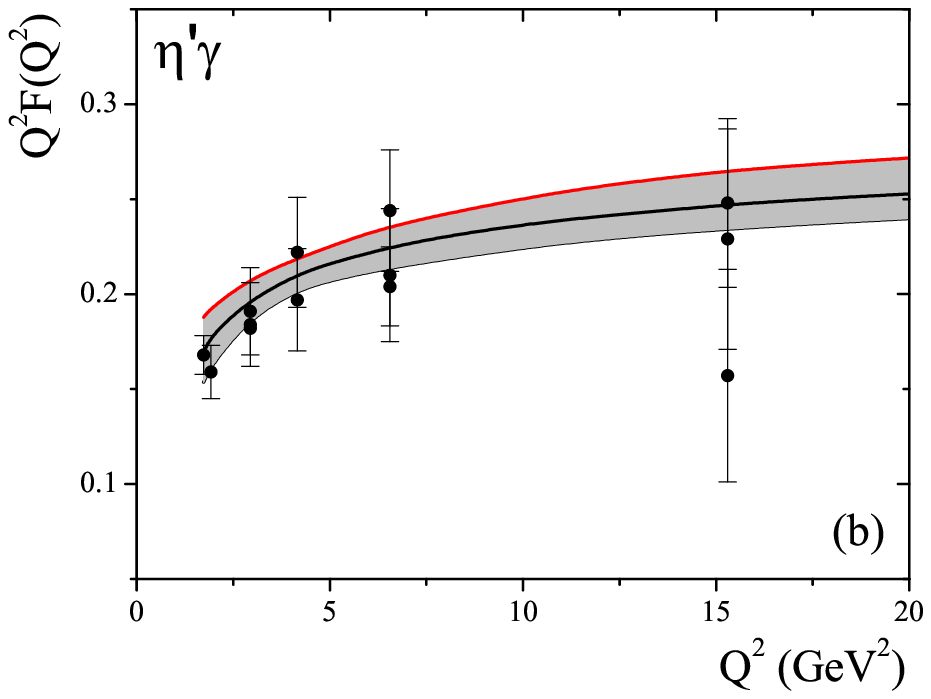,height=7.9cm,width=7.1cm,clip=}
\vspace{0.0cm} \caption{ The $\eta\gamma$ (left) and
$\eta^{\prime}\gamma$ (right) scaled transition form factors as
functions of $Q^2$. The central solid curves are found using the
values $B_2^q(\eta_1)=B_2^q(\eta_8)=0.05$ and $B_2^g(\eta_1)=17$.
The shaded areas demonstrate $1\sigma$ regions for the transition
FFs.} \label{fig:Gfig3}
\end{figure}

\begin{figure}[t]
\centering\epsfig{file=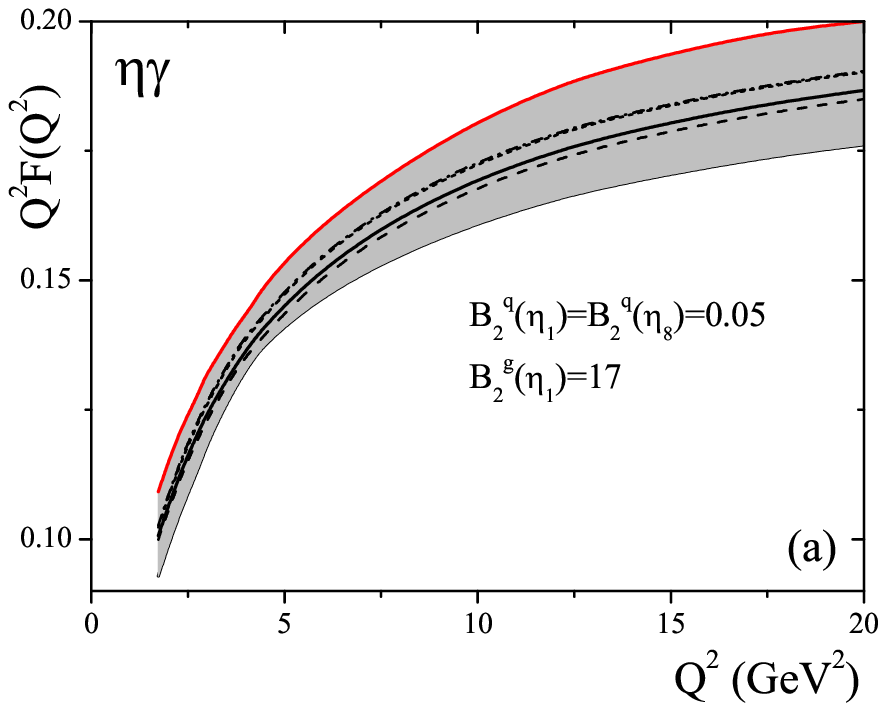,height=8cm,width=7.0cm,clip=}
\hspace{1.5cm}
\centering\epsfig{file=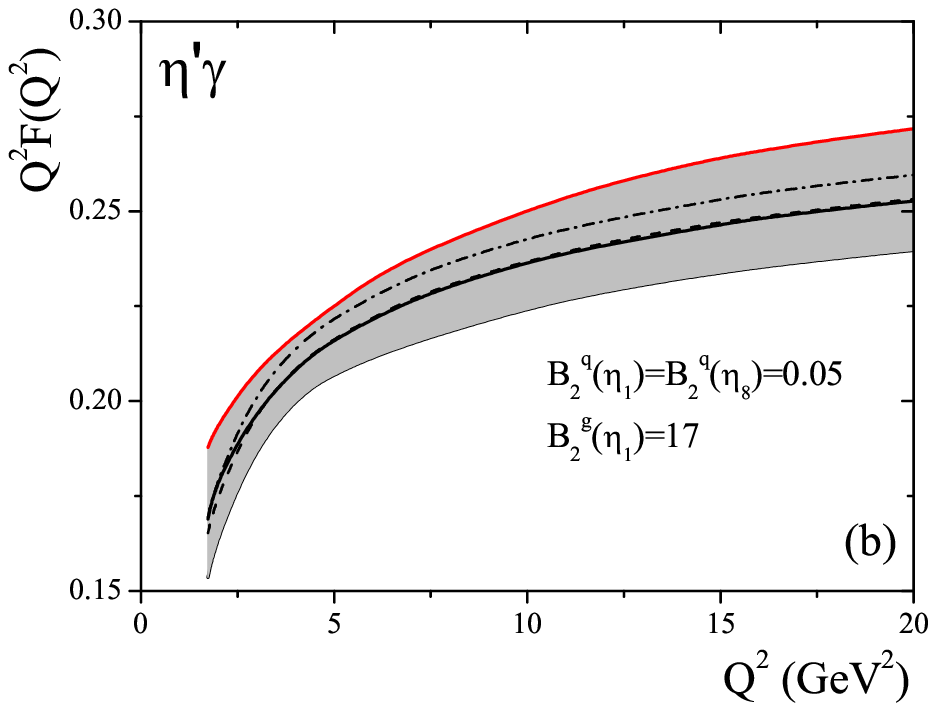,height=8.0cm,width=7.1cm,clip=}
\vspace{0.0cm} \caption{ The dependence of the $\eta\gamma$ (left
panel) and $\eta^{\prime}\gamma$ (right panel) scaled transition
form factors on the values of the decay constants $f_1$ and $f_8$.
The octet-singlet mixing angle is $\theta_p=-15.4^{\circ}$. The
solid curves in both panels are calculated using
$f_1/f_{\pi}=1.17$, $f_8/f_{\pi}=1.26$. The long-dashed curves
correspond to $f_1/f_{\pi}=1.17$, $f_8/f_{\pi}=1.24$ and the
short-dashed one in the left panel are found by employing the
values $f_1/f_{\pi}=1.17$, $f_8/f_{\pi}=1.30$. The dash-dotted
curves in both panels describe FFs\ obtained with
$f_1/f_{\pi}=1.20$ and $f_8/f_{\pi}=1.28$.} \label{fig:Gfig4}
\end{figure}

\begin{figure}[t]
\centering\epsfig{file=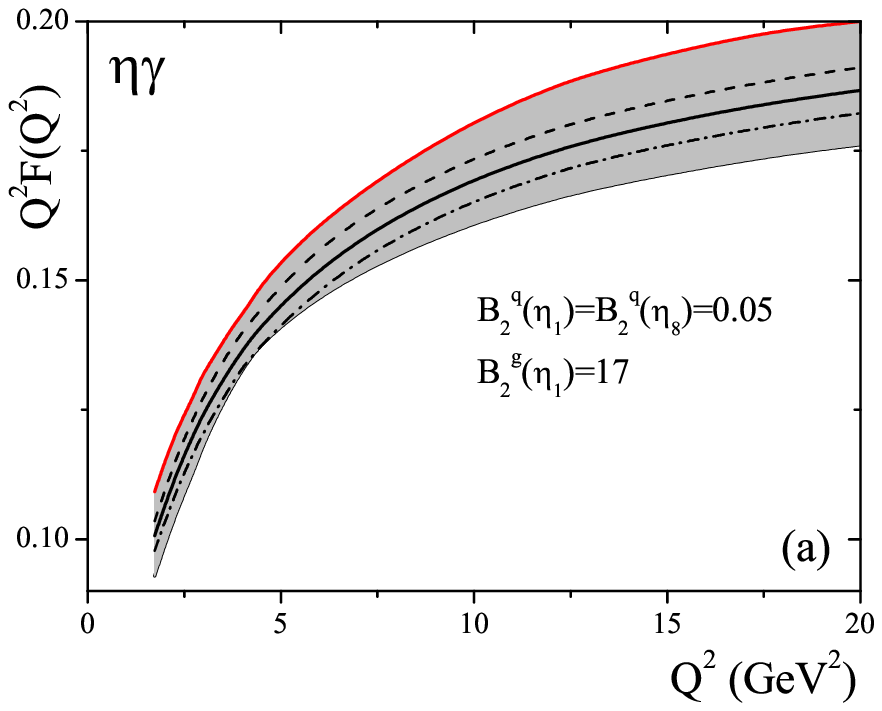,height=8cm,width=7.0cm,clip=}
\hspace{1.5cm}
\centering\epsfig{file=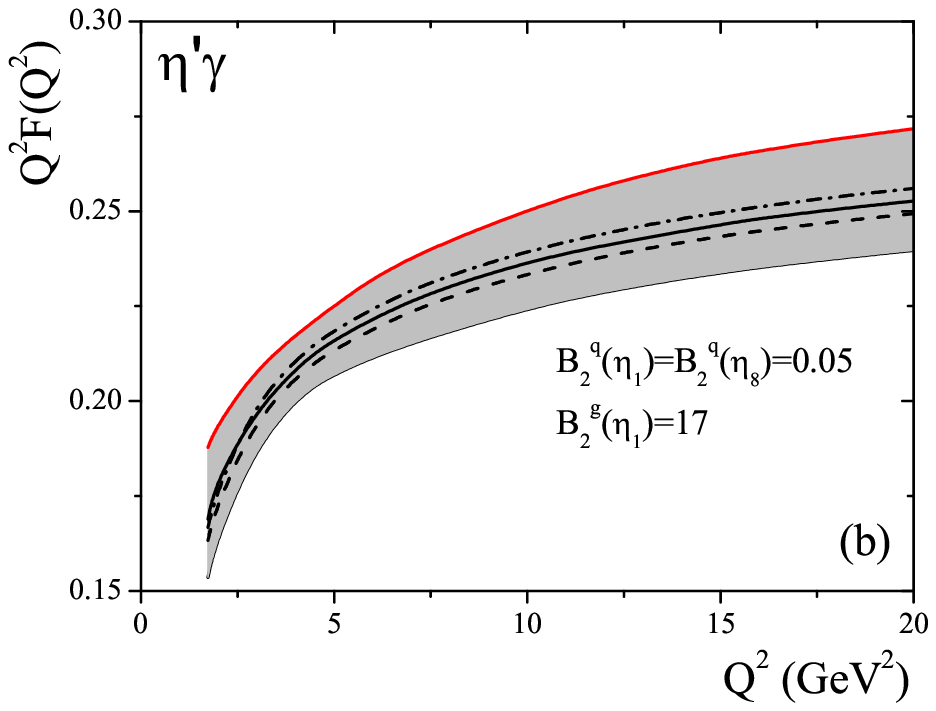,height=8.0cm,width=7.1cm,clip=}
\vspace{0.0cm} \caption{The dependence of the $\eta\gamma$ (left
panel) and $\eta^{\prime}\gamma$ (right panel) scaled transition
form factors on the octet-singlet mixing angle $\theta_p$.
The solid curves describe the ``default'' choice
$\theta=-15.4^{\circ}$.
The dashed curves correspond to
$\theta=-16.4^{\circ}$ and the dash-dotted ones to
$\theta=-14.4^{\circ}$.} \label{fig:Gfig5}
\end{figure}

\begin{figure}[t]
\centering\epsfig{file=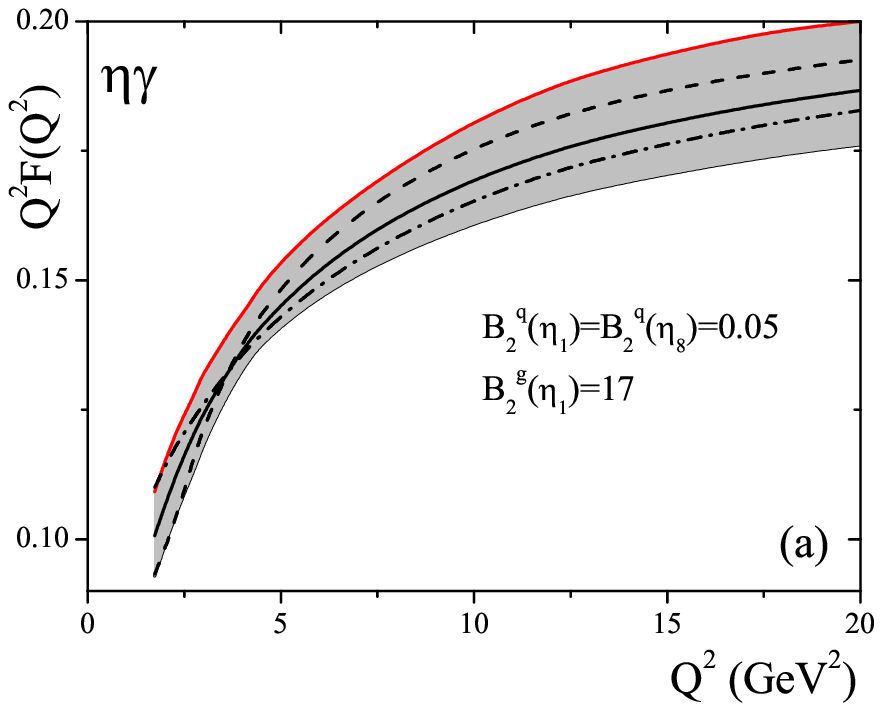,height=8cm,width=7.0cm,clip=}
\hspace{1.5cm}
\centering\epsfig{file=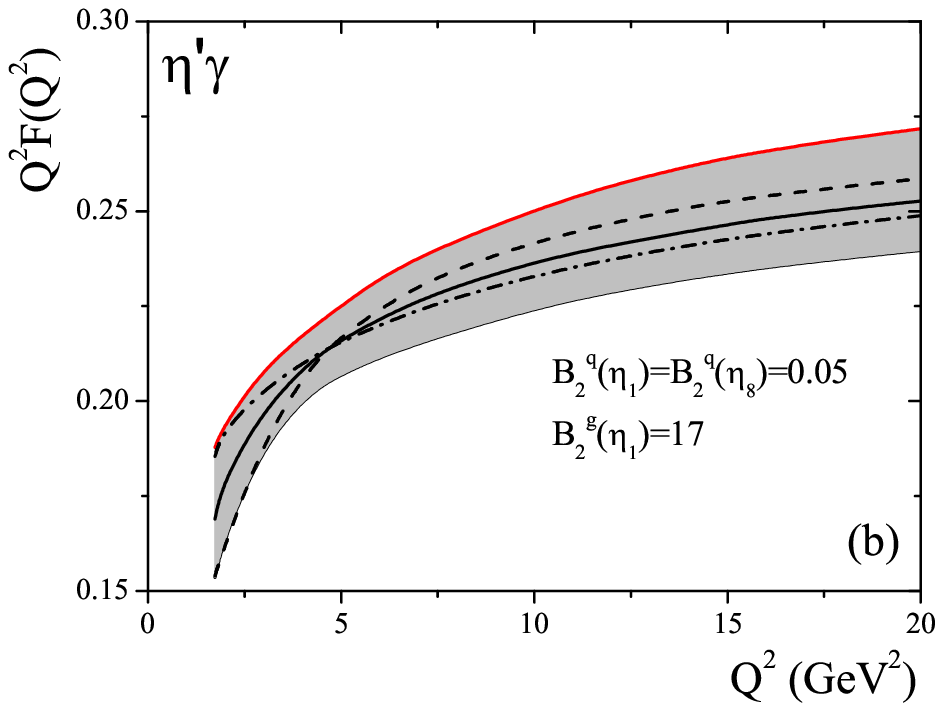,height=8.1cm,width=7.1cm,clip=}
\vspace{0.0cm} \caption{The $\eta\gamma$ (left panel) and
$\eta^{\prime}\gamma$ (right panel) scaled transition form factors
as functions of $Q^2$.
All predictions have been obtained within the ordinary mixing scheme
and using the initial input parameters (\ref{eq:12}) and
(\ref{eq:50}).
The broken lines denote the FFs\ with the
uncertainties included via Eq.\ (\ref{eq:42d}), and using the
following values of $\{N_q\}, \; q=1,2,3,4$ : 0.9
(dashed lines); -0.6 (dash-dotted lines).} \label{fig:Gfig6}
\end{figure}

\begin{figure}[t]
\centering\epsfig{file=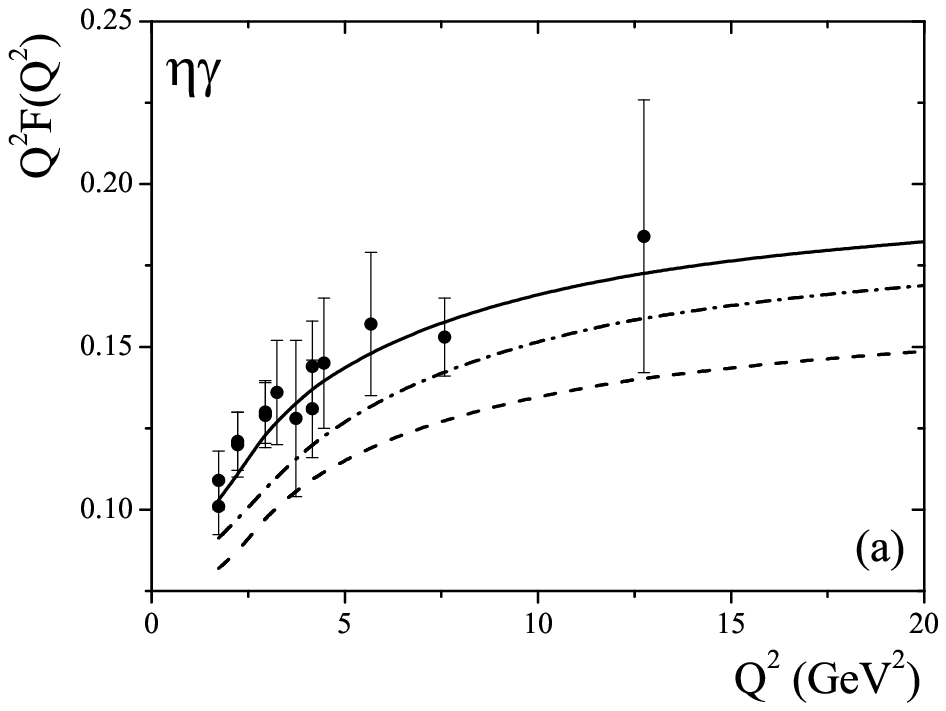,height=8.1cm,width=7.0cm,clip=}
\hspace{1.5cm}
\centering\epsfig{file=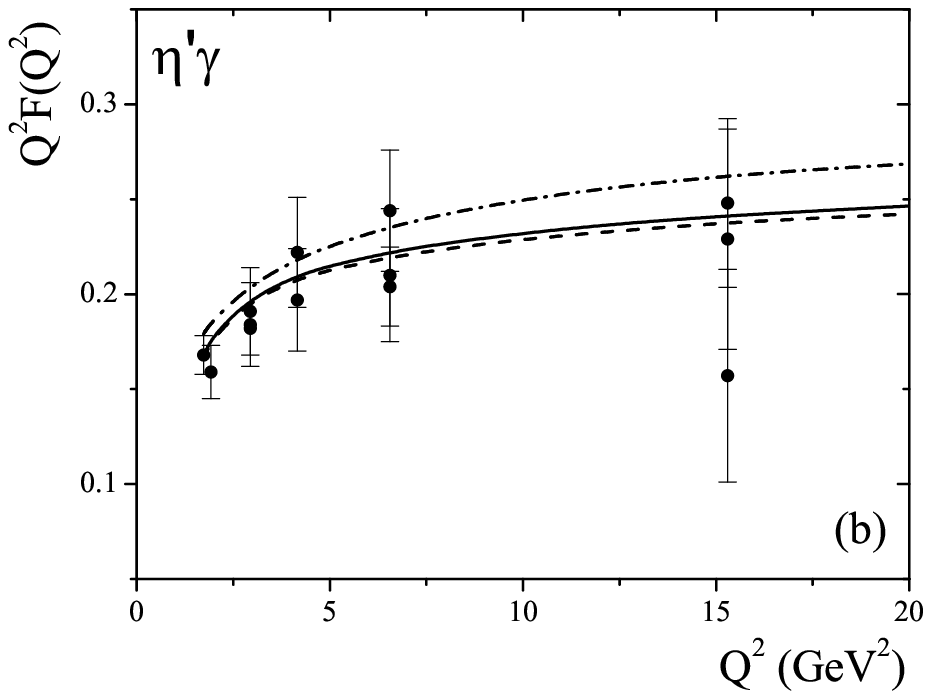,height=7.9cm,width=6.7cm,clip=}
\vspace{0.0cm} \caption{The $\eta\gamma$ (a) and
$\eta^{\prime}\gamma$ (b) electromagnetic transition FFs\ vs.\
$Q^2$.
The solid lines correspond to the ordinary octet-singlet mixing
scheme with parameters $B_2^q(\eta_1)=B_2^q(\eta_8)=0.02$
and $B_2^g(\eta_1)=18$.
The broken lines are obtained within the two-mixing angle scheme.
The dashed lines describe the situation with the same parameters as
the solid curves.
The parameters for the dash-dotted curves are
$B_2^q(\eta_1)=B_2^q(\eta_8)
=0.15,\,B_2^g(\eta_1)=18$.}
\label{fig:Gfig7}
\end{figure}

\begin{figure}[t]
\centering\epsfig{file=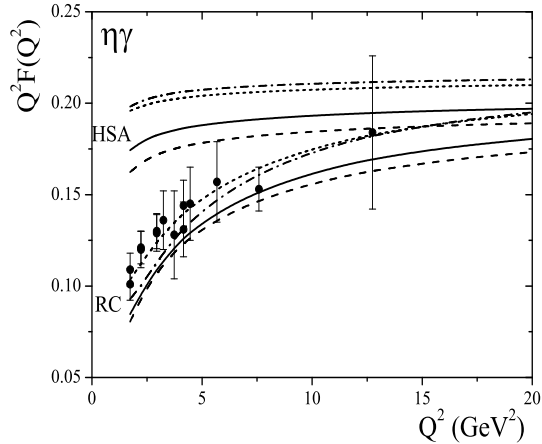,height=7cm,width=8.0cm,clip=}
\hspace{1.5cm} \vspace{0.0cm} \caption{The $\eta \gamma $ scaled
transition FF\ vs. $Q^2$.
In the  computations the ordinary octet-singlet
mixing scheme is used.
The upper (lower) bundle of curves is found within the standard HSA\
(RC\ method).
The correspondence between the curves and the input parameters is as
follows: for the solid curves
$B_2^q(\eta_1)=B_2^q(\eta_8)=0,\,B_2^g(\eta_1)=0$; for the dashed
lines $B_2^q(\eta_1)=B_2^q(\eta_8)=-0.05,\,B_2^g(\eta _1)=0$; for
the dash-dotted ones $B_2^q(\eta_1)=B_2^q(\eta_8)=0.1,\,B_2^g(\eta
_1)=0$, and for the short-dashed curves
$B_2^q(\eta_1)=B_2^q(\eta_8)=0.1,\,B_2^g(\eta_1)=15$.}
\label{fig:Gfig8}
\end{figure}

\begin{figure}[t]
\centering\epsfig{file=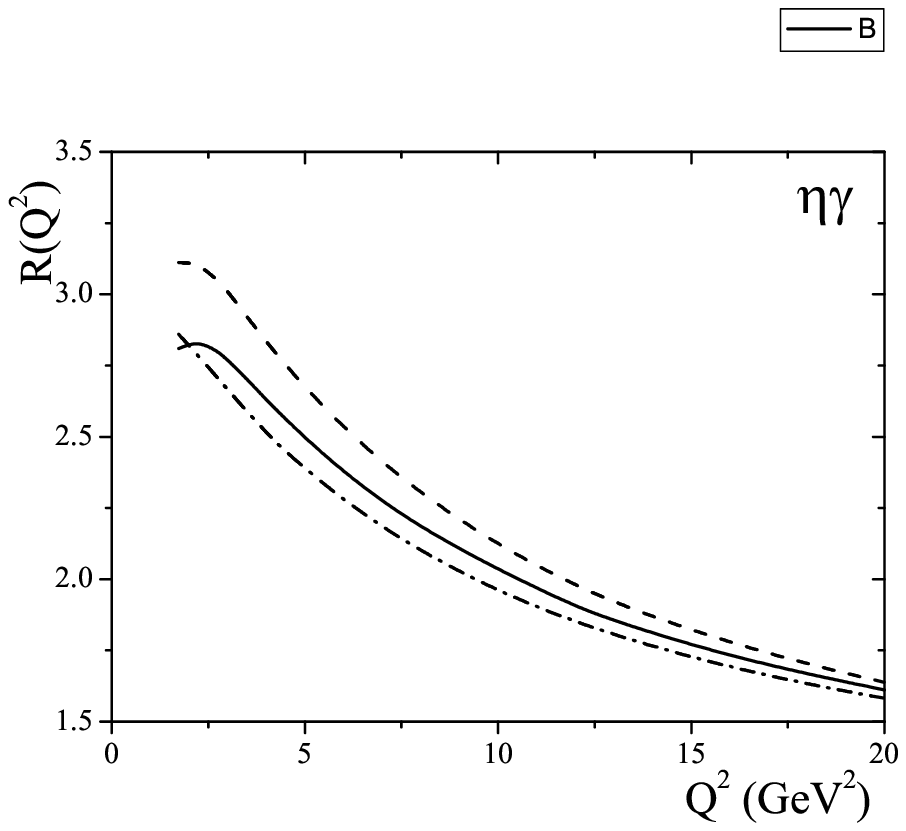,height=7cm,width=8.0cm,clip=}
\hspace{1.5cm} \vspace{0.0cm} \caption{The ratio $R(Q^2)$ for the
$\eta \gamma $ FF.
The solid line corresponds to the input parameters
$B_2^q(\eta _1)=B_2^q(\eta _8)=B_2^g(\eta _1)=0$.
The dash-dotted curve describes the same ratio, but for $B_2^q(\eta
_1)=B_2^q(\eta _8)=0,\,B_2^g(\eta _1)=14$, while the dashed one
corresponds to
$B_2^q(\eta _1)=B_2^q(\eta _8)=0.05,\,B_2^g(\eta_1)=10$.}
\label{fig:Gfig9}
\end{figure}

The results shown in Figs.\ \ref{fig:Gfig1} --
\ref{fig:Gfig9}---with the exception of Fig.\
\ref{fig:Gfig7}---are obtained within the ordinary octet-singlet
mixing scheme. In Fig.\ \ref{fig:Gfig1} the predictions for the
$\eta\gamma$ and $\eta^{\prime}\gamma$ FFs\ are presented for
$B_2^q(\eta_1)=B_2^q(\eta_8)=0.02$ and various values of
$B_2^g(\eta_1)$. One appreciates that without the gluon
contribution ($B_2^g(\eta_1)=0$) both FFs\ are slightly below the
data points, especially $Q^2F_{\eta^{\prime}\gamma}(Q^2)$. But
their deviations are not dramatic and to improve the agreement
with the data, one has to include the contribution coming from the
gluon component of the $\eta_1\gamma$ transition FF. The
corresponding results are shown in Fig.\ \ref{fig:Gfig1} by broken
lines. These numerical calculations demonstrate that the gluonic
contribution is important
at relatively low values of the momentum-transfer $Q^2$. From
Fig.\ \ref{fig:Gfig1} it is clear that the gluonic contribution,
arising from the $\eta_1$ DA with $B_2^g(\eta_1)>0$, enhances the
transition FFs\ $Q^2F_{\eta \gamma}(Q^2)$ and
$Q^2F_{\eta^{\prime}\gamma}(Q^2)$ in the region $1.5\;{\rm
GeV}^2\leq Q^2\leq 12\;{\rm GeV}^2$ while reducing their magnitude
at $Q^2>12\;{\rm GeV}^2$. This effect is sizeable for the
$\eta^{\prime}\gamma$ transition FF relative to its counterpart
for $\eta \gamma$, in particular, for larger values of
$B_2^g(\eta_1)$ and for smaller values of $B_2^g(\eta _1)$. The
impact of the gluonic contribution on the $\eta\gamma,\;
\eta^{\prime}\gamma$ transition FFs\ is quite understandable,
recalling that the physical $\eta$ and $\eta^{\prime}$ states
consist predominantly of the flavor $SU_f(3)$ octet $\eta_8$ and
singlet $\eta_1$ states, respectively, with the $\eta_1\gamma$
transition FF comprising also a gluonic part. Therefore, the
$\eta^{\prime}\gamma$ transition FF should be and is more
sensitive to the gluonic part.

These features of the $\eta\gamma$ and $\eta^{\prime}\gamma$
transition FFs\ determine the $1\sigma$ region for the allowed
values of the Gegenbauer coefficients
$B_2^q(\eta_1)=B_2^q(\eta_8)$ and $B_2^g(\eta_1)$, plotted
in Fig.\ \ref{fig:Gfig2}.
In other words, the $\eta\gamma$ and $\eta^{\prime}\gamma$
transition FFs, computed in the context of the RC\ method by
employing the model DAs\ with input parameters belonging to the
shaded region in Fig.\ \ref{fig:Gfig2}, describe the CLEO data
with a $1\sigma$ accuracy.

In Fig.\ \ref{fig:Gfig3} we plot the $1\sigma$ areas for the
$\eta\gamma$ and $\eta^{\prime}\gamma$ transition FFs.
If we were to consider the $\eta\gamma$ and $\eta^{\prime}\gamma$
transitions separately, these areas would be larger than those shown
in Fig.\ \ref{fig:Gfig3}.
For the $\eta\gamma$ transition, the upper bound of the $1\sigma$
region can be extended towards larger values of
$Q^2F_{\eta\gamma}(Q^2)$.
For the $\eta^{\prime}\gamma$ transition, the lower bound of the
corresponding $1\sigma$ region can be shifted towards lower values
of $Q^2F_{\eta^{\prime}\gamma}(Q^2)$.
But their joint treatment leads to the picture
drawn in Fig.\ \ref{fig:Gfig3}.

A major problem in extracting the values of theoretical parameters
from the experimental data is their stability against
uncertainties inherent in the theoretical expressions. In the case
under consideration, expressions (\ref{eq:4}), (\ref{eq:35}), and
(\ref{eq:36}) for the $\eta\gamma$ and $\eta^{\prime}\gamma$
transition FFs\ depend on the factorization scale, on the QCD
scale parameter $\Lambda$, the decay constants $f_1,\; f_8$, and
the octet-singlet mixing angle $\theta_p$. As we have explained in
Sec.\ \ref{sec:borel}, the renormalization scale $\mu_{\rm R}^2$
($\overline{\mu}_{\rm R}^2$) in the context of the RC\ method is
determined by the hard-scattering dynamics of the underlying
partonic subprocess and is not a free parameter. Our analytical
expressions for the transition FFs, calculated by keeping
$\mu_{\rm F}^2 \neq Q^2$, allow us to analyze the dependence of
the extracted parameters $B_2^q(\eta_1),\;B_2^q(\eta_8)$, and
$B_2^g(\eta_1)$ on the factorization scale $\mu_{\rm F}^2$. We
have performed the computation of the $\eta\gamma$ and
$\eta^{\prime}\gamma$ transition FFs\ using the values $\mu_{\rm
F}^2=Q^2/2$ and $\mu_{\rm F}^2=2Q^2$ and found that our prediction
for the $1\sigma$ area (Fig.\ \ref{fig:Gfig2}) is absolutely
stable against these variations. This means that the FFs
determined by the input parameters from the $1\sigma$ area in
Fig.\ \ref{fig:Gfig2}, by varying the factorization scale, remain
within the corresponding $1\sigma$ regions shown in Fig.\
\ref{fig:Gfig3}. Stated differently, the variation of $\mu_{\rm
F}^2$ in the limits $\mu_{\rm F}^2 \in [1/2,\ 2]Q^2$ does not
change (shift, rotate) the $1\sigma$ area in Fig.\
\ref{fig:Gfig2}. On the contrary, the variation of the QCD scale
parameter $\Lambda$ modifies the $1\sigma$ region in Fig.\
\ref{fig:Gfig2}. The entailed modifications shift the region along
both axes, retaining, however, its form stable. Thus, computations
performed with $\Lambda_4=0.26\ {\rm GeV}$ result in the following
shifts: along the $B_2^q(\eta_1)=B_2^q(\eta_8)$ axis: $\sim
0.005$, along the $B_2^g(\eta_1)$ axis: $\sim 1$. Hence, the
modification of the $1\sigma$ area is $\sim 9 \%$ in the first and
$\sim6\%$ in the second direction, respectively, the percentages
being given relative to the central values (see, Eq.\
(\ref{eq:54a}) below).

The response of the central curves (Fig.\ \ref{fig:Gfig3}) on
variations of the decay constants $f_1,\;f_8$, and such due to the
octet-singlet mixing angle $\theta_p$ within corresponding
phenomenologically allowed ranges \cite{FKS98}, is demonstrated in
Figs.\ \ref{fig:Gfig4} and \ref{fig:Gfig5}. It is remarkable that
under these variations the central curves remain entirely within
the associated $1\sigma$ areas for the $\eta\gamma$ and
$\eta^{\prime}\gamma$ transition FFs. It turns out that the
$\eta\gamma$ transition FF is more sensitive to the value of the
decay constant $f_8$ than the $\eta^{\prime}\gamma$ one. The
results for the $\eta^{\prime}\gamma$ transition FF\ obtained by
varying the constant $f_8 \in [1.24,\;1.30]f_{\pi}$ at fixed
$f_1=1.17f_{\pi}$ practically coincide with each
other.\footnote{This is the reason why in Fig.\ \ref{fig:Gfig4}(b)
the FF corresponding to the values
$f_1=1.17f_{\pi},\;f_8=1.30f_{\pi}$ is not displayed.} On the
contrary, the $\eta^{\prime}\gamma$ transition FF\ demonstrates a
rather strong dependence on the decay constant $f_1$, whereas the
$\eta\gamma$ one is stable under such variations (cf.\ the
short-dashed and dash-dotted curves, respectively, in Fig.\
\ref{fig:Gfig4}(a)). Our computations with $\theta_p=-15.4^{\circ}
\pm 1^{\circ}$ confirm the conclusion drawn in Ref.\ \cite{Aga01}
that the FF for the $\eta\gamma$ transition is more sensitive to
$\theta_p$ than the one for the $\eta^{\prime}\gamma$ transition.
Summing up, we can state that the modification of the central
curves in Figs.\ \ref{fig:Gfig4} and \ref{fig:Gfig5}, due to the
variations of the decay constants and the mixing angle discussed
above, does not exceed the level of $\pm 3 \%$ of their values.

In Sec.\ \ref{sec:borel} we have emphasized that the ambiguities
produced by the principal value prescription, inherent in the RC\
method, affect the predictions for the transition FFs\ in
accordance with Eq. (\ref{eq:42d}). The ambiguity $[Q^2F_{\rm
M\gamma }^{1(8)}(Q^2)]^{\rm amb}$ depends on the $\eta_1$ and
$\eta_8$ DAs\ and also on the constant $\{N_q\}$. In reality,
however, for given DAs\ of the $\eta_1$ and $\eta_8$ states, the
available experimental information allows one to extract
constraints on $\{N_q\}$. To effect the influence of such
contributions, we show exemplarily in Fig.\ \ref{fig:Gfig6}
predictions for the FFs\ with and without such ambiguities,
utilizing the expansion coefficients
$B_2^q(\eta_1)=B_2^q(\eta_8)=0.05,\,B_2^g(\eta_1)=17$. We find
that in order that the FFs\ remain within the corresponding
$1\sigma$ regions, the upper and lower bounds, respectively, for
the constants $\{N_q\}$ are provided by the values $\{N_q=0.9\}$
and $\{N_q=-0.6\}$. Hence, the $\eta\gamma$ transition FFs\ with
the ambiguities included, corresponding to $\{N_q=-0.6\}$
($\{N_q=0.9\}$) at $Q^2<4~{\rm GeV}^2$, are larger (smaller) than
the FFs\ without such corrections and are, in addition, smaller
(larger) for $Q^2>4~{\rm GeV}^2$. For the $\eta^{\prime}\gamma$
transition FF\ we observe, qualitatively, the same behavior, but
with $Q^2 \simeq 5~ {\rm GeV}^2$ as the transition momentum scale
from the small to the large (and vice versa) regions. In any case,
the uncertainties do not exceed the level of $\pm 11\%$ of the
corresponding FFs\ in the region $Q^2\sim (1.73-2)\,\,{\rm GeV}^2$
and reach a mere $\mp 3\%$ level in the region $Q^2\sim
(16-20)\,\,{\rm GeV}^2$.

Below, we present sample estimates for the eigenfunctions expansion
coefficients of the $\eta_1$ and $\eta_8$ DAs\ in the context of the
ordinary mixing scheme:
\begin{equation}
\label{eq:51}
B_2^q(\eta_1)=0\, ,\; B_2^q(\eta_8)=0\, , \;
B_2^g(\eta_1) \in [17,\,21.5]\, ,
\end{equation}
\begin{equation}
\label{eq:51a}
B_2^q(\eta_1)=0.02\, ,\; B_2^q(\eta_8)=0.02\, , \;
B_2^g(\eta_1)\in [16,\,20.5]\, ,
\end{equation}
\begin{equation}
\label{eq:52}
B_2^q(\eta_1)=0.05\, ,\; B_2^q(\eta_8)=0.05\, , \;
B_2^g(\eta_1) \in [15,\,19]\, ,
\end{equation}
and
\begin{equation}
\label{eq:53}
B_2^q(\eta_1)=0.1\, ,\; B_2^q(\eta_8)=0.1\, , \;
B_2^g(\eta_1) \in [13.5,\,17]\, .
\end{equation}
The constraints (\ref{eq:51})--(\ref{eq:53}) on the input
parameter $B_2^g(\eta_1)$ are extracted for fixed
coefficients $B_2^q(\eta_1)$ and $B_2^q(\eta_8)$, and represent
the $1\sigma$ range for the values of $B_2^g(\eta_1)$ compatible
with the CLEO data.
Restrictions on the parameters $B_2^q(\eta_1)$ and
$B_2^q(\eta_8)$ at fixed value of $B_2^g(\eta_1)$ can also be
derived.
For example, for $B_2^g(\eta_1)=16$, we get
\begin{equation}
\label{eq:54}
B_2^g(\eta_1)=16,\;B_2^q(\eta_1)=B_2^q(\eta_8) \in
[0.02,\, 0.11]\, .
\end{equation}
Summarizing this point, the estimates for the Gegenbauer coefficients
$B_2^q(\eta_1),\,B_2^q(\eta_8)$ and $B_2^g(\eta_8)$ in the DAs for the
$\eta_1,\;\eta_8$ states are
\begin{equation}
\label{eq:54a}
B_2^q(\eta_1)=B_2^q(\eta_8)
=0.055 \pm 0.065,\;\;B_2^g(\eta_1)=18 \mp 4.5.
\end{equation}

Here some comments concerning the usual octet-singlet mixing
scheme (\ref{eq:11}) and the parameter set (\ref{eq:12}) are in
order. These parameters were extracted from the analysis of the
CLEO data using the two-mixing-angles scheme, but staying within
the context of the hard-scattering approach of perturbative QCD.
Our computations demonstrate that adopting the RC method, the
parameters given by (\ref{eq:12}) satisfactorily describe these
data, provided one uses the usual octet-singlet mixing scheme.
Therefore, one can consider the parameters (\ref{eq:12}) as a
\emph{prediction} of the RC method and the one-angle mixing
scheme. This prediction differs from those obtained already within
the one-angle mixing scheme, but employing the traditional
theoretical methods (see, for example, Ref.\ \cite{ONS})

However, our calculations do not exclude the usage of the
two-mixing-angle scheme in conjunction with the RC method. But in
such a case, a considerably larger contribution of the
nonasymptotic terms to the DAs of the $\eta_1$ and $\eta_8$ states
would be required. Carrying out such a computation via
(\ref{eq:13}), (\ref{eq:14}), we obtained the results shown in
Fig.\ \ref{fig:Gfig7}. Inspection of Fig.\ \ref{fig:Gfig7}(a)
reveals that the $\eta\gamma$ transition FF\ found within this
scheme lies significantly lower than the data. Therefore, to
improve the results, a relatively large contribution of the first
Gegenbauer polynomial to the DAs\ of the $\eta_1$ and $\eta_8$
states seems necessary. In Fig.\ \ref{fig:Gfig7} we display the
FFs\ obtained using the parameters
$B_2^q(\eta_1)=0.15,\;B_2^q(\eta_8)=0.15$ and $B_2^g(\eta_1)=18$.
We consider the values $B_2^q(\eta_1)=B_2^q(\eta_8)=0.15$ as
determining the lower bound for the admissible set of DAs\ in the
context of the two-mixing-angles parametrization scheme. Hence, in
the two-mixing-angles scheme, we obtain
\begin{equation}
\label{eq:55}
B_2^q(\eta_1)=B_2^q(\eta_8)=0.15,\;B_2^g(\eta_1) \in
[16,\,20]\, .
\end{equation}

The $\eta$ and $\eta^{\prime}$ DAs\ were extracted from the CLEO
data on the $\eta \gamma$ and $\eta^{\prime}\gamma$ transition
FFs\ \cite{KP02} having also recourse to the $\eta^{\prime}$-meson
energy spectrum in the decay $\Upsilon(1S)\to \eta^{\prime}X$
\cite{AP03}. In both these papers the standard HSA\ was employed.
In Ref.\ \cite{KP02}, estimates for the parameters
$B_2^1(\mu_0^2),\;B_2^g(\mu_0^2)$, and $B_2^8(\mu_0^2)$ were made
within the two-mixing-angles scheme (\ref{eq:13}), reading
\begin{equation}
\label{eq:56}
B_2^1(1\,\,\,{\rm GeV}^2) = -0.08\pm 0.04\, ,\;
B_2^g(1\,\,{\rm GeV}^2) = 9\pm 12\, ,\; B_2^8(1\,\,{\rm GeV}^2) =
-0.04\pm 0.04\, .
\end{equation}
These coefficients are related to ours through the
expressions
\begin{equation}
\label{eq:57} B_2^1(\mu_0^2)=\frac{A(\mu_0^2)}6\, ,\;
B_2^g(\mu_0^2) =\frac{B(\mu_0^2)}5\, , \;
B_2^8(\mu_0^2)=\frac{C(\mu_0^2)}6\, .
\end{equation}
Using our approach and the one-angle mixing scheme,
the values of these parameters were determined to be
$$
B_2^1(1\,\,{\rm GeV}^2)=-0.214\pm 0.025\, ,\; B_2^g(1\,\,{\rm
GeV}^2) =19.25\pm 2.25\, ,\; B_2^8(1\,\,{\rm GeV}^2)=0\, ,
$$
$$
B_2^1(1\,\,{\rm GeV}^2)=-0.183\pm 0.025\, ,\; B_2^g(1\,\,{\rm
GeV}^2) =18.57\pm 2.25\, ,\; B_2^8(1\,\,{\rm GeV}^2)=0.02\, ,
$$
\begin{equation}
\label{eq:58} B_2^1(1\,\,{\rm GeV}^2)=-0.139\pm 0.022\, ,\;
B_2^g(1\,\,{\rm GeV}^2) =17.8\pm 2\, ,\; B_2^8(1\,\,{\rm
GeV}^2)=0.05\, ,
\end{equation}
$$
B_2^1(1\,\,{\rm GeV}^2)=-0.07\pm 0.02\, ,\; B_2^g(1\,\,{\rm
GeV}^2) =16.85\pm 1.75\, ,\;B_2^8(1\,\,{\rm GeV}^2)=0.1\, ,
$$
and
\begin{equation}
\label{eq:59} B_2^1(1\,\,{\rm GeV}^2)=-0.1128\pm 0.045\, , \;
B_2^g(1\,\,{\rm GeV}^2) =16.208\pm 0.144\, ,\; B_2^8(1\,\,{\rm
GeV}^2)=0.65 \pm 0.045\, .
\end{equation}

In the case of the two-mixing-angles scheme, we find
\begin{equation}
\label{eq:60}
B_2^1(1\,\,{\rm GeV}^2)=-0.050\pm 0.022\, ,\; B_2^g(1\,\,{\rm GeV}%
^2)=20.4\pm 2\, ,\; B_2^8(1\,\,{\rm GeV}^2)=0.15\, .
\end{equation}
One observes that within the two-mixing-angles scheme, the
parameters $B_2^1,\,B_2^8$ obey the constraints $B_2^1(1\, {\rm
GeV}^2)<0$ and $B_2^8(1\, {\rm GeV}^2)>0$ (cf.\ Eq.\
(\ref{eq:56})).

On the other hand, the constraints for the parameters
$B_2^q(\mu_0^2)$ and $B_2^g(\mu_0^2)$, extracted in Ref.\
\cite{AP03} at the normalization point $\mu_0^2=2\,\,{\rm GeV}^2$,
read
\begin{equation}
\label{eq:61}B_2^q(2\;{\rm GeV}^2)=0.010\pm 0.068,\quad
B_2^g(2\;{\rm GeV}%
^2)=5.6\pm 3.4\, .
\end{equation}
Comparing now Eq.\ (\ref{eq:61}) with the values given in Eq.\
(\ref{eq:55}), and taking into account that in Ref.\ \cite{AP03}
different values for the scheme parameters defined by Eq.\
(\ref{eq:50}) were used, we come to the conclusion that in the
context of the RC\ method and the two-mixing-angles scheme, the
region $B_2^q(2\ {\rm GeV}^2)<0$ should be excluded as
contradicting the CLEO data.

The apparent discrepancy between the results of the present work
and those of Ref.\ \cite{KP02}, as regards the extracted values of
the coefficients $B_2^1(1\;{\rm GeV}^2)$ and $B_2^8(1\;{\rm
GeV}^2)$, is related to the fact that the employed theoretical
schemes are intrinsically different. Indeed, the transition FFs\
computed in the standard HSA\ overshoot the CLEO data---especially
in the low-momentum transfer regime. In Fig.\ \ref{fig:Gfig8} the
$\eta\gamma$ transition FF obtained in the standard HSA\ and the
ordinary octet-singlet mixing scheme is depicted. One appreciates
that the deviation from the data is considerable. The DAs\
corresponding to the parameters $B_2^q(\eta_1),\;B_2^q(\eta _8)>0$
even increase this disagreement, whereas by adding the gluon
component with $B_2^g(\eta _1)>0$ one can reduce it. Therefore to
decrease the magnitude of the FFs, and achieve this way a better
agreement with the data, the standard HSA\ would call for the
two-mixing-angles scheme and for DAs\ mainly with
$B_2^q(\eta_1),\,B_2^q(\eta_8)<0$. The inclusion of power
corrections changes this situation radically. In fact, at
low-momentum transfer these corrections enhance the absolute value
of the NLO correction to the FFs\ by more than a factor of $2.5-3$
and, because the contribution of the NLO term to the FFs\ is
negative, power corrections reduce the leading-order prediction
for the FFs\ considerably, while at the highest $Q^2$ values
measured by the CLEO collaboration this influence becomes more
moderate. As a result, the $\eta\gamma$ and $\eta^{\prime}\gamma$
transition FFs\ computed using the input parameters from the
$1\sigma$ area in Fig.\ \ref{fig:Gfig2} within the RC\ method in
conjunction with the one-angle mixing scheme are in agreement with
the CLEO data. In order to quantify these statements, we show in
Fig.\ \ref{fig:Gfig9} the numerical results for the ratio
\begin{equation}
\label{eq:62}
R_{\rm M\gamma}(Q^2)
=\frac{[Q^2F_{\rm M\gamma}(Q^2)]_{\rm NLO}^{\rm res}}{%
[Q^2F_{\rm M\gamma }(Q^2)]_{\rm NLO}^{\rm HSA}}
\end{equation}
for some selected values of the expansion coefficients.

\section{Concluding remarks}

In this work we have performed a computation of the $\eta \gamma$
and $\eta^{\prime}\gamma$ transition FFs\ within the RC\ method.
The latter has enabled us to estimate a class of power corrections
to the FFs\ related to nonperturbative effects arising from the
dependence of the strong coupling on the longitudinal momentum
fractions of the partons inside the $\eta$ and $\eta^{\prime}$
mesons after the identification of the renormalization scale with
a physical momentum depending on these fractions. This has been
achieved by regularizing the infrared singularities ensuing from
the end points $x=0,1$ by means of the principal value
prescription within the IR\ renormalon approach. The effect of
power-suppressed ambiguities to the considered form factors was
addressed and their influence was found to be less important,
though not negligible, with contributions varying in the range
between $3\%$ at high to $11\%$ at low $Q^2$ values.

Contributions to the FFs\ from the valence quark as well as the
two-gluon Fock-state of the $\eta$ and $\eta^{\prime}$ meson DAs\
have been taken into account.
We have obtained the Borel resummed expressions
$[Q^2F_{\rm M\gamma}(Q^2)]^{\rm res}$ for the FFs\ and proved that
in the asymptotic limit $Q^2 \to \infty$ they lead to the standard
HSA\ predictions.

We have demonstrated that the effect of the calculated power
corrections on the $\eta\gamma$ and $\eta^{\prime}\gamma$
transition FFs\ is considerable. Indeed, at moderate values of the
momentum-transfer $Q^2 \leq 5 \; {\rm GeV}^2$ they turn out to
enhance the absolute value of the $O(\alpha_{{\rm s}})$ correction
to the FFs\ more than $2.5-3$ times. The ratio $R_{\rm
M\gamma}(Q^2)$ of the corresponding contributions depends on the
specific $M\gamma$ transition under consideration and on the input
parameters (Gegenbauer coefficients) of the $\eta$ and
$\eta^{\prime}$ meson DAs. These features of the power corrections
have important consequences: the enhanced (negative) NLO
correction significantly reduces the leading-order contribution to
the FFs, so that the input parameters of the $\eta$ and
$\eta^{\prime}$ meson DAs, which correctly describe the CLEO data
within the RC\ method, must obey the constraints presented in
Fig.\ \ref{fig:Gfig2} by the shaded $1\sigma$ area to fulfill Eq.\
(\ref{eq:54a}). It is worth emphasizing that our predictions for
the $\eta$ and $\eta^{\prime}$ meson DAs\ disagree with those
extracted from the CLEO data in the context of the standard HSA.

The DAs\ of the $\eta$ and $\eta^{\prime}$ mesons obtained in this
work can be useful in the investigation of other exclusive
processes that involve $\eta$ and $\eta^{\prime}$ mesons,
especially at lower momentum-transfer values, where the standard
HSA\ is most unreliable.

\end{document}